\documentclass[sigconf]{acmart} 
\newcommand{\etal}{et~al.~}

\usepackage{color,soul}
\usepackage{fancybox} % Import the fancybox package

\newcommand{\promptbox}[1]{
  \begin{center} % Center the box
    \doublebox{%
      \begin{minipage}{.95\columnwidth} % Control the width of the content inside the box
        \vspace{5pt} % Add some vertical space at the top inside the box
        #1
        \vspace{5pt} % Add some vertical space at the bottom inside the box
      \end{minipage}%
    }
  \end{center}
}

\AtBeginDocument{%
  \providecommand\BibTeX{{%
    \normalfont B\kern-0.5em{\scshape i\kern-0.25em b}\kern-0.8em\TeX}}}

\newcommand{\cc}[1]{{\color{black} #1}}

\copyrightyear{2024}
\acmYear{2024}
\setcopyright{acmlicensed}\acmConference[CHI '24]{Proceedings of the CHI Conference on Human Factors in Computing Systems}{May 11--16, 2024}{Honolulu, HI, USA}
\acmBooktitle{Proceedings of the CHI Conference on Human Factors in Computing Systems (CHI '24), May 11--16, 2024, Honolulu, HI, USA}
\acmDOI{10.1145/3613904.3642363}
\acmISBN{979-8-4007-0330-0/24/05}

\begin{document}

\title[An Empathy-Based Sandbox Approach to Bridge the Privacy Gap]{An Empathy-Based Sandbox Approach to Bridge \cc{the Privacy Gap among Attitudes, Goals, Knowledge, and Behaviors}}

\author{Chaoran Chen}
\email{cchen25@nd.edu}
\affiliation{%
  \institution{University of Notre Dame}
  \city{Notre Dame}
  \state{Indiana}
  \country{USA}
}

\author{Weijun Li}
\authornote{Work done as a visiting student at the University of Notre Dame.}
\email{mvs@zju.edu.cn}
\affiliation{%
  \institution{Zhejiang University}
  \city{Hangzhou}
  \country{China}
}

\author{Wenxin Song}
\authornotemark[1]
\email{wsong23@nd.edu}
\affiliation{%
  \institution{The Chinese University of Hong Kong, Shenzhen}
  \city{Shenzhen}
  \country{China}
}

\author{Yanfang Ye}
\email{yye7@nd.edu}
\affiliation{%
  \institution{University of Notre Dame}
  \city{Notre Dame}
  \state{Indiana}
  \country{USA}
}

\author{Yaxing Yao}
\email{yaxing@vt.edu}
\affiliation{%
  \institution{Virginia Tech}
  \city{Blacksburg}
  \state{Virginia}
  \country{USA}
}

\author{Toby Jia-Jun Li}
\email{toby.j.li@nd.edu}
\affiliation{%
  \institution{University of Notre Dame}
  \city{Notre Dame}
  \state{Indiana}
  \country{USA}
}

\renewcommand{\shortauthors}{Chen, et al.}

\begin{abstract}
\end{abstract}





%     \centering
%     \includegraphics[width=0.8\textwidth]{figures/teaser_paper.png}
%      \caption{An empathy-based approach where users interact with online services with different personas in a risk-free sandbox without leaking their real personal data. Users can observe and experience the causal effect between their privacy configurations/behaviors and system outcomes, acquire privacy knowledge, and translate the knowledge into actual behavior.}
%     \label{fig: teaser}


\begin{abstract}
Managing privacy to reach privacy goals is challenging, as evidenced by the privacy attitude-behavior gap. Mitigating this discrepancy requires solutions that account for both system opaqueness and users' hesitations in testing different privacy settings due to fears of unintended data exposure. We introduce an empathy-based approach that allows users to experience how privacy attributes may alter system outcomes in a risk-free sandbox environment from the perspective of artificially generated personas. To generate realistic personas, we introduce a novel pipeline that augments the outputs of large language models (e.g., GPT-4) using few-shot learning, contextualization, and chain of thoughts. Our empirical studies demonstrated the adequate quality of generated personas and highlighted the changes in privacy-related applications (e.g., online advertising) caused by different personas. Furthermore, users demonstrated cognitive and emotional empathy towards the personas when interacting with our sandbox. We offered design implications for downstream applications in improving user privacy literacy.
% the feasibility of eliciting realistic responses from privacy-related applications (e.g., online advertising) using personas. 

\end{abstract}
%The privacy paradox is a long-standing issue in the privacy domain, referring to the discrepancy between users' attitudes and their actual behaviors in managing their privacy.
%However, mitigating this issue presents challenges from both system and user perspectives. System opaqueness hinders users' informed decision-making, while users' fear of exposing personal data further discourages them from understanding the influence of the privacy data. To address these challenges, we propose an empathy-based method that allows users to observe the correlation between their privacy choices and system outcomes in a risk-free environment. Through the approach, users can observe target ads using the identities of personas generated by large language models. We use few-shot learning, contextualization, and chain of thoughts to enhance their authenticity. Our empirical studies demonstrated that the generated personas have higher information consistency than baseline GPT-generated personas and revealed users’ empathy toward the generated personas in the context of target ads.
\maketitle

\section{Introduction}
Managing privacy to achieve individuals' privacy goals is challenging~\cite{acquisti2020secrets}, as evidenced by the discrepancy between people's privacy attitudes and their actual behaviors \cite{norberg2007privacy, kokolakis2017privacy}. Such inconsistency has been observed in various domains, such as social networking service \cite{hallam2017online, wu2019privacy, obar2020biggest}, online shopping \cite{beresford2012unwillingness, anic2019determinants, bandara2020explicating}, mobile app \cite{pentina2016exploring, barth2019putting, rowe2020contact} and Internet-of-Things \cite{williams2016perfect}. 

Bridging the \cc{gap among privacy attitudes, goals, knowledge, and behavior can be difficult for two reasons.} 
From a system perspective, the inherent \textit{opaqueness} in the system prevents users from making informed decisions about protecting their privacy. The asymmetric information \cite{acquisti2015privacy} provided by the system makes it difficult for users to understand what data is collected and how other parties use it \cite{mazzia2012pviz, yao2017folk, li_privacystream_2017}. Consequently, users are unable to make informed decisions to safeguard their personal data while maintaining the desired level of usability and system utility, such as whether to opt out of certain data collection practices, configure the frequency and granularity of data sharing, or the adoption of privacy-enhancing tools.
From the user perspective, the \textit{fear of exposing personal data} \cite{zrenner2019usage, spiekermann2019data, li_privacy_2020} while navigating an opaque system can further discourage users from experimenting with different possible privacy configurations to link their available options of privacy choices to their consequences, thereby reinforcing the system's opaqueness. Once users share their private data, they will no longer have control over how the other party utilizes the information \cite{pasquale2015black}. In addition, another barrier that prevents users from meeting their privacy goals is their \textit{limited experience and lack of privacy literacy} \cc{\cite{solove2021myth}}. Lay users are prone to perceive fewer privacy threats compared to technicians \cite{kang2015my}. Thus, even if users sometimes know their privacy goals, they still trade off their privacy for convenience, as they believe that their data are well protected by the system.

% add one short paragraph for the previous approach: privacy education and nudge. and what is the shortcoming of them/
To support users' privacy decision-making, two approaches have been widely applied: privacy education and nudging \cite{wisniewski2017making}. Privacy education endeavors to cultivate privacy awareness and literacy, thereby equipping users with the knowledge to make well-informed privacy choices. However, a notable challenge with these methods lies in the extended duration required for shifts in privacy attitudes. Users often encounter difficulties in adhering to expert privacy recommendations and translating acquired knowledge into specific online contexts. An alternative approach to facilitating privacy decisions is through privacy nudging\cite{acquisti2009nudging}. Nudges encompass subtle yet influential prompts that steer individuals toward certain behaviors. Although nudges can facilitate the adoption of specific behaviors, their effects tend to be transitory, as intermittent adjustments in individuals' privacy practices may not necessarily extend to their overall privacy literacy.

Motivated by the aforementioned challenges and limitations of current approaches, we present an \textit{empathy-based} method that allows users to experience and observe the correlation between their privacy data and the system outcomes in a real-time and risk-free environment.
In this approach, we use \textit{personas} that come with synthesized personal data based on real-world \cc{system outcomes}. Each persona represents a fictional user \cite{blomkvist2002persona} with a distinctive biography, demographic information, and a large set of synthesized personal data. For example, here is an exemplary biography of a ``tangible'' persona: 
\begin{itemize}
    \item Alice is a 40-year-old white woman living in New York. She is an administrative assistant, and her annual income is around ninety thousand USD. She lives with her husband and two teenage children. She is an avid user of social media platforms such as Facebook and Instagram, where she often shares posts, photos, and videos of her life. She also often purchases clothing items and books on Amazon. 
\end{itemize}
Unlike personas often used in the user experience design process, personas used in our context should also include plausible \textit{realistic} longitudinal personal data such as web browsing history, social media logs, location records, and weekly schedules. 
%interaction logs with (fictional) friends and family members, and online purchase history. 
The intricate realness of these personas is facilitated by the use of Large Language Models (LLMs), which can generate a diverse range of highly detailed and modifiable personas. Our design draws upon the principles of empathy-based design. Recognized for its essential role in user experience and persuasive design, the empathy-based design employs narrative and role-play techniques to establish deeper and more meaningful connections with users \cite{cuff2016empathy, wright2008empathy, buchenau2000experience}.

Through the sandbox, users can interact with different online services, as usual, using the identities of their selected personas. The sandbox will be loaded with personal data from the persona instead of the user, so whenever an online service queries personal data, the synthetic personal data associated with the persona will be provided.
As far as service providers are concerned, the data appear real, causing them to offer personalized content and services as though interacting with the user the persona represents.
This gives users a risk-free platform to investigate privacy settings and actions, perceive the resulting user experience, notice the tangible consequences of their privacy choices, and experience emotional results, positive or negative, in a convincingly interactive environment without exposing their actual personal data.

% \tlcomment{need 1--2 paragraph summarizing the study and its findings}

We validate the proposed approach through a prototype (i.e., Privacy Sandbox) and a study involving 15 participants. The results validated the technical feasibility of our approach to generate artificial personas with realistic synthesized personal data. Our findings imply that users can indeed establish empathy with personas when using the Privacy Sandbox and identify links between the persona's privacy attributes with the observed system outcomes. 
The results of the study also offer design implications for using the proposed approach to empower users to acquire privacy knowledge.
\cc{Contributions of this work represent a first step towards empowering users to empathize with generated artificial personas, understand their own privacy goals, and gain privacy knowledge. Our broader motivation is to use acquired privacy knowledge to encourage changes in privacy behavior, ultimately bridging the gap between attitudes, goals, knowledge, and actual behavior.}

This paper makes the following contributions:
\begin{itemize}
    \item Introduces an empathy-based approach that allows users to experience the links between privacy \cc{attributes} and system outcomes in a risk-free sandbox environment using artificially generated personas.
    
    \item  Validates the viability of the proposed approach through a proof-of-concept implementation and empirical studies. The study results confirmed the users' cognitive and emotional empathy toward the generated personas when interacting with the sandbox in the context of target advertisements.

    \item Discusses the design implications of adopting this empathy-based privacy persona approach to empower users to acquire privacy knowledge that leads to behavior change in the future.
\end{itemize}

\section{Related work}
\label{sec:related_work}

\subsection{Empirical studies on the privacy attitudes-behavior gap}
% Empirical studies on privacy paradox
% tune down the privacy paradox and focus more on the "gap"

% measure the gap

Most empirical studies measure the gap between users' privacy attitudes and behaviors, commonly using surveys. For example, Madejski \etal \cite{madejski2012study} used surveys to gauge privacy attitudes, previous privacy settings, and self-reported sharing intentions on Facebook, identifying potential sharing violations by comparing intentions with settings. Colnago \etal \cite{colnago2023there} also employed within-subjects surveys, revealing mismatches between attitudes/preferences and behaviors. 
Although surveys are able to explore privacy attitudes, they are not reliable when examining irregular or infrequent privacy behavior \cite{staddon2013self, kokolakis2017privacy}. Consequently, many studies combine surveys with experiments to collect more reliable behavior data. For instance, Norberg \etal \cite{norberg2007privacy} assessed willingness to disclose information in surveys and later conducted a field study to compare the willingness with actual disclosure, finding significant differences. Barth \etal \cite{barth2019putting} measured privacy concerns through surveys and compared the results with the actual behavior of the participants, represented by the number of downloaded intrusive apps.

Although multiple previous studies have measured the privacy attitude-behavior gap, only a few have proposed ways to address it. Previous research has examined risk awareness~\cite{dienlin2015privacy}, the privacy calculus ~\cite{park2013digital}, and digital nudges~\cite{ioannou2021privacy, story2022increasing} as potential solutions.
Sutanto \etal \cite{sutanto2013addressing} designed a personalized privacy-safe application that retains user information locally on their smartphones while still providing them with personalized products.
Mattson \etal ~\cite{mattson2023close} suggested changing negative attitudes in different functional areas to reduce the intention-behavior gap. 

% , with numerous studies observing~\cite{reynolds2011sharing,strahilevitz2016privacy,barth2019putting} and interpreting~\cite{acquisti2004privacy, acquisti2005privacy,lee2013people,baek2014my,gerber2018explaining} this phenomenon. 

\cc{The inconsistency between privacy attitudes and behaviors has been framed by some researchers as the ``privacy paradox''~\cite{norberg2007privacy}. As summarized by Solove~\cite{solove2021myth}, research supporting the privacy paradox argues that actual privacy behaviors better indicate individual's true privacy preferences compared to self-reported privacy attitudes. This privacy paradox results from the distortion of privacy behavior.
However, opponents either argue that this is not a privacy issue or that the privacy paradox does not exist.
Some studies diverted the phenomenon from the privacy domain and regarded it as a trust issue~\cite{lutz2014privacy, alabdali2021privacy}. 
Other researchers refuted the concept of the privacy paradox. For example, Martin~\cite{martin2020breaking} found that consumers keep strong privacy expectations even after the disclosure of information. Solove~\cite{solove2021myth} argued that the behaviors in privacy paradox studies pertain to specific contexts, while the stated privacy concerns are much more general.
To explain this mixed view, Acquisti et al.~\cite{acquisti2020secrets} explained that it is due to researchers' different definitions and interpretations of the ``paradox''.
Despite the controversy over the term, the evidence supporting the gap between the mental state of privacy and actual behavior is strong~\cite{acquisti2006imagined,madejski2012study,adjerid2018beyond,barth2019putting}, which is the main motivation of this work.
}

% \tlcomment{I think we still have to survey/discussion the privacy paradox here. Just need to make it clear that it's a controversial topic (with citations and discussions on why), but the existence of the gap (which is our main motivation) is evidence and widely accepted.}

Unlike previous methods relying on surveys and/or experiments for on-the-spot decisions, our approach offers a risk-free environment for users to experience online services by using different generated personas' data. This allows users to reflect on system outcomes and make more informed decisions, potentially mitigating the privacy \cc{attitudes-behavior} gap resulting from ad-hoc decisions.

% \tlcomment{can we find more work that tried to tackle this issue? i found it hard to believe that there is only one paper in this space}
% \mvscomment{fixed}
%% game - transfer knowledge to behavior, behavior change
%Gamification presents an alternative method to encourage users to translate their knowledge of privacy into actions. Park \etal \cite{park2009role} provided some role-play situations (\eg web game, telephone phishing) for students and asked them to act in the roles and think of how to protect their privacy. Digital-PASS \cite{ghazinour2020novel, ghazinour2019digital} is another cyber attack-defense simulation game that allows users to act as posters in social media and hackers. When acting as posters, users only accumulate online audiences but also protect their accounts by regularly changing their passwords and identifying phishing scams. While as hackers, they can use the post library to conduct simulated theft actions on the posters. Through this role-playing game, users can experience a previously unavailable perspective on social media and identify more privacy threats.

% argument:
% Unlike previous work, which usually asked people to make on-spot decisions through surveys or experiments, our method takes a step back as we create a risk-free sandbox for users to explore the outcomes of their different privacy configurations or behaviors and reflect on different system outcomes. Therefore, users can think twice before making decisions, which we hope can mitigate the gap in privacy paradox that derives from their ad-hoc decision-making. 

\subsection{Approaches to enhance privacy awareness and behavior}
%Approaches to prompt privacy behavior change
Previous theories on privacy behavior~\cite{ajzen1991theory, bandura1977social,larose2007promoting} emphasize the role of attitudes in shaping behavior. 
Based on them, we distinguish two approaches in \cc{improving privacy knowledge and }promoting privacy behavior change: the \textit{top-down approach}, which primarily seeks to foster privacy attitudes and literacy to indirectly influence behavior, and the \textit{bottom-up approach}, which directly influences behavior through techniques like nudging. We aim to comprehensively assess their effectiveness and improve their impact on privacy behavior.

Top-down approaches (e.g., privacy education) aim to improve privacy behavior by fostering privacy awareness~\cite{10.1145/3600096,10.1145/3411764.3445516} and enhancing privacy literacy~\cite{10.1145/3576050.3576153}. 
For instance, Desimpelaere et al.~\cite{desimpelaere2020knowledge} observed that privacy literacy training improves children's understanding and promotes privacy-protective behavior. Sideri et al.~\cite{sideri2019enhancing} found that university-based education enhanced students' digital knowledge and privacy awareness on social networking sites.
Innovative methods like Franco et al.~\cite{franco2023fostering} employed technology-enhanced pedagogical scenarios to involve students in active learning by using their own social media traces.
Despite increased awareness and knowledge, top-down approaches may not consistently translate into behavior change. Users often struggle to apply experts' privacy advice~\cite{das2019typology} due to its vagueness~\cite{reeder2017152} and lack of alignment with their specific needs and contexts~\cite{zou2020examining}.

Bottom-up approaches (e.g., digital nudging) alter user privacy behavior by guiding their choices in digital environments using techniques such as visual cues~\cite{10.1145/3213765}, information presentation~\cite{carpenter2018designing}, default settings~\cite{baek2014changing}, and incentives~\cite{an2021intention, barbosa2019if}. These nudges can reduce data disclosure and influence privacy choices in a short period~\cite{carpenter2018designing, eling2016investigating}. 
% Hansen and Jespersen~\cite{hansen2013nudge} categorize nudges with varied transparency, which have different privacy behavior impacts~\cite{zhang2016privacy}.
While privacy nudges can alter privacy behavior quickly, they often lead to temporary effects as they may not necessarily improve users' long-term privacy literacy or decision alignment with their attitudes.

Our approach combines both top-down and bottom-up approaches. It offers a systematic risk-free platform for users to learn about privacy in a structured way (top-down) while providing experiential learning~\cite{gentry1990experiential} through real-world consequences based on user interactions and privacy choices (bottom-up). This dual approach aims to bridge the gap between privacy knowledge and behavior effectively.

% Our work combines both top-down and bottom-up approaches. On one side, our method is systematic, it embodies the top-down approach to privacy, offering users a platform to systematically explore a risk-free interface over a period of time,  helping them learn about privacy in a structured way. On the other side, our approach is also experiential. It integrates the bottom-up ethos by enabling users to interact, experiment, and understand the real-world consequences of their privacy choices and settings based on their own experiences. 

% education - beyond privacy 
\subsection{The use of empathy in user experience design and persuasive design}
% % privacy, persuasive design, learning
Empathy, often defined as the capacity for an affective response aligned with someone else's situation rather than one's own~\cite{batson1987prosocial}, encompasses both affective and cognitive components. \textit{Affective empathy} involves an immediate emotional response to others, while \textit{cognitive empathy} pertains to understanding others' feelings \cite{feshbach1975empathy, kouprie2009framework, cuff2016empathy, drouet2022empathy}. It is a powerful instrument to connect people with others and has been applied in various domains like user experience (UX) design~\cite{cuff2016empathy} and persuasive design.

% Empathy is commonly defined as ``an affective response appropriate to someone else's situation rather than one’s own.''~\cite{batson1987prosocial}
% It encompasses both affective and cognitive components: \textit{affective empathy} involves an immediate emotional response to others, and \textit{cognitive empathy} is to understand other's feelings \cite{feshbach1975empathy, kouprie2009framework, cuff2016empathy, drouet2022empathy}. It is a powerful instrument to connect people with others, and it has been applied in many areas such as user experience design \cite{cuff2016empathy} and persuasive design. 

In UX field, empathic design \cite{kouprie2009framework} aims to enable designers to ``step into the user's shoes'' and ``walk the user's walk'', thereby crafting products that align with user needs. A solid foundation for comprehending empathy in design research is established through the exploration of the literature on philosophy, psychology, and neuroscience. Surma-Aho et al. ~\cite{surma2022conceptualization} offered a comprehensive review of empathy's role in design. A pertinent framework, proposed by Hess and Fila~\cite{hess2016manifestation}, defines empathy along two axes: affective experiences vs. cognitive processes and self-oriented vs. other-oriented perspectives, yielding four dimensions: 
\begin{enumerate}
    \item \textit{perspective-taking}, where designers imagine users' thoughts and feelings (cognitive, other-oriented);
    \item \textit{empathic concern}, as designers display sincere care for users (affective, other-oriented);
    \item \textit{emotional congruence}, with designers sharing users' emotional states (affective, self-oriented);
    \item \textit{projection}, when designers experience unease due to users' challenges (cognitive, self-oriented).
\end{enumerate}
Building on it, various approaches, such as narrative and role-play techniques~\cite{wright2008empathy}, have been developed to foster deeper connections with users and their experiences in empathic design. These methods involve creating scenarios and personas to envision potential design innovations~\cite{carroll1997scenario} or simulating user experiences through role-play~\cite{buchenau2000experience}.

Our work draws inspiration from these frameworks and design methods of empathic understanding, with the aim of investigating whether users can develop empathy toward the generated personas and whether this influences their acquisition of privacy knowledge.

In persuasive design, the stimulation of empathy is a crucial technique~\cite{caraban201923}. Previous studies have used the malleable nature of empathy to promote prosocial behavior ~\cite{crowley2011behavioral,sierksma2015group,taylor2019accountability}. For example, Taylor et al.~\cite{taylor2019accountability} found that embedding empathy nudges in social media posts can encourage bystander intervention for cyberbullying victims. Additionally, many researchers have found that designing with empathy can encourage the natural empathetic behavior of people who have existing social ties or shared interests~\cite{taylor2019accountability,preece1999empathic} A typical example is that VR can enhance cognitive empathy by emphasizing user similarities~\cite{shin2018empathy}.

Our motivation is aligned with the concept of bystander empathy~\cite{taylor2019accountability}, aiming to modify user behavior by eliciting their empathy towards generated personas. 
To achieve this goal, we draw upon empathy-inducing techniques from persuasive design, such as providing detailed and specific information~\cite{crowley2011behavioral,small2003helping}, immersive role-playing and perspective-taking~\cite{shin2018empathy,belman2010designing}, and considering connections between users and personas~\cite{shin2018empathy}.

Although there are various measurement scales to quantify empathy in psychology~\cite{baron2009essential, davis1980multidimensional, hogan1969development, escalas2003sympathy}, our focus is on the relationship between users and personas rather than personal characteristics. We derive inspiration from approaches used to measure designers' empathy during the design process, including indicators such as empathic expressions, personal experiences, respectful questioning, and discussing user facts~\cite{van2011achieving}. Our empathy measurement approach combines self-report methods and integrates established theoretical frameworks of empathy.

\section{An empathy-based privacy persona approach}

\subsection{Overview}
\label{sec:approach_overview}
% \tlcomment{highlevel: I feel this section still intertwine two different things: the high-level approach and our implementation of the approach}

% \tlcomment{Figure 1 -- the green color is too light. I recommend (1) use black color only; (2) only bold the keywords e.g., emphathize, acquire privacy knowledge, and experience (instead of observe)}

% \yaxing{This sentence talks about two things: 1) empathy-based approach; 2) risk-free sandbox environment. They seem to mix the two concepts together. We might need to introduce them separately, starting with the empathy-based approach (high-level idea), then the sandbox (our implementation)}

% \yaxing{we also need to explain how the process of building empathy works in this context, and how does that help uses to understand more about the system outcomes/privacy consequences, etc.}

In an effort to bridge the gap between users' attitudes and their behaviors in managing their privacy, we introduce a new empathy-based sandbox approach. This approach uses artificially generated user personas with realistic synthesized personal data, enabling users to (1) load synthetic personal data into browsers; (2) interact with websites and applications ``under the disguise'' of an artificial persona; (3) experiment with various privacy configurations and behaviors in a risk-free environment; and (4) experience the corresponding outcomes (both positive and negative).

% \tlcomment{get that old figure back here}
\begin{figure*}
    \centering
    \includegraphics[width=0.9\linewidth]{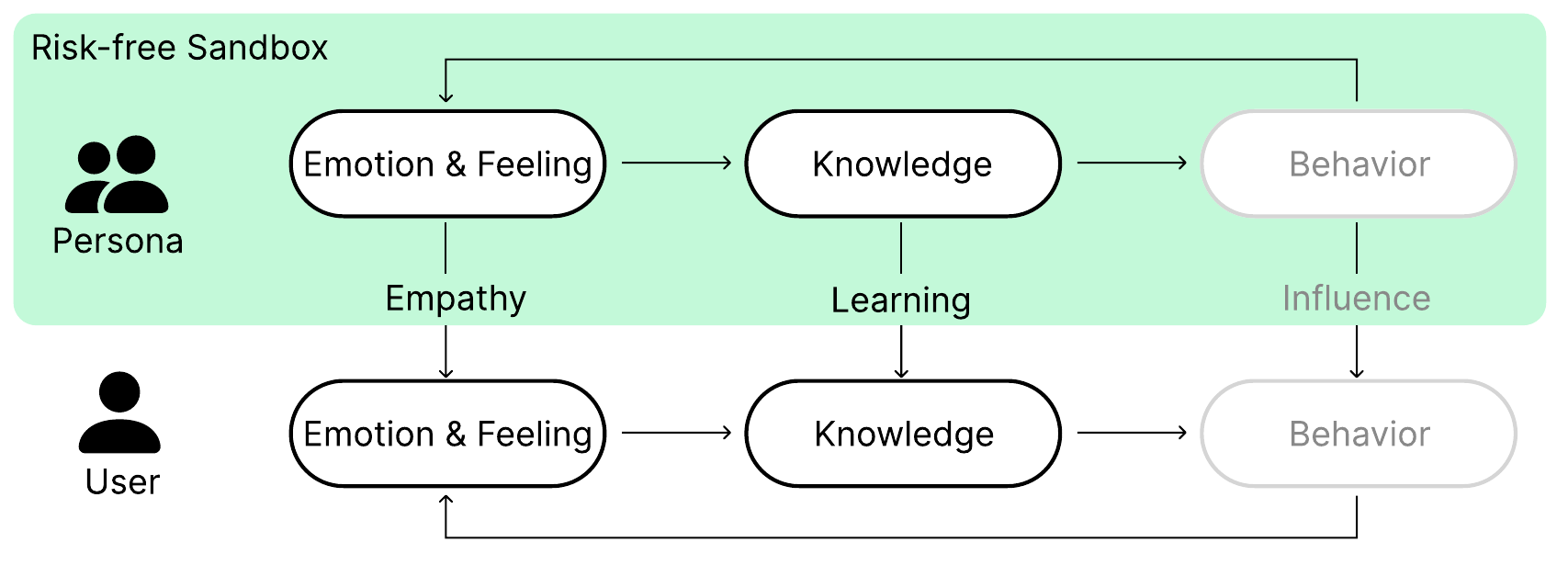}
    \caption{An empathy-based approach where users interact with online services with different personas in a risk-free sandbox without leaking their real personal data. Users can observe and experience the causal effect between their privacy configurations/behaviors and system outcomes, acquire privacy knowledge, and translate the knowledge into actual behavior. \cc{Note that this is a general framework. We leave the studies of changes in privacy behaviors as future work.}}
    \label{fig:teaser}
\end{figure*}

As illustrated in Fig. \ref{fig:teaser}, the key goals of our empathy mechanisms are two-fold. 

\begin{enumerate}
    \item \textbf{Emotional Resonance}: When users encounter \cc{system outcomes based on the personal data} when acting as personas, we believe that users can feel the emotion that the persona would have felt (e.g., frustration or anger from privacy violations; joy from apt personalized recommendations). Users will also realize how they may feel when they encounter a similar incident.
    \item \textbf{Knowledge Acquisition}: When users experience an outcome of their privacy behaviors (e.g., seeing a particular personalized ad or being influenced by an algorithmic decision) when acting as personas, we believe that users will be able to identify patterns and acquire generalizable knowledge, which fosters a more intuitive understanding of likely outcomes from their future actions.
\end{enumerate}

As discussed in Section \ref{sec:related_work}, it is feasible to stimulate users' empathy toward privacy personas by employing empathy-inducing techniques from user experience design~\cite{wright2008empathy} and persuasive design~\cite{taylor2019accountability}. Simultaneously, users can learn about privacy in a structured and interactive manner by experiencing the influence of privacy personas' information on system outcomes. Such experiential learning not only provides real-time feedback, akin to nudging techniques~\cite{hansen2013nudge} but also facilitates users to acquire privacy knowledge~\cite{10.1145/3576050.3576153}.
Therefore, combining emotional resonance with privacy knowledge acquisition can result in more \textit{motivated} and \textit{informed} users. Rooted in current frameworks and past research, we hypothesize that fostering both emotional resonance and knowledge acquisition can promote privacy literacy, subsequently leading to changes in user privacy behaviors that align with their preferences.

% We proposed an empathy-based approach to allow users to experience how the system's access to personal data can influence system outcomes in a risk-free sandbox environment. 

\begin{figure*}
    \centering
    \includegraphics[width=0.9\linewidth]{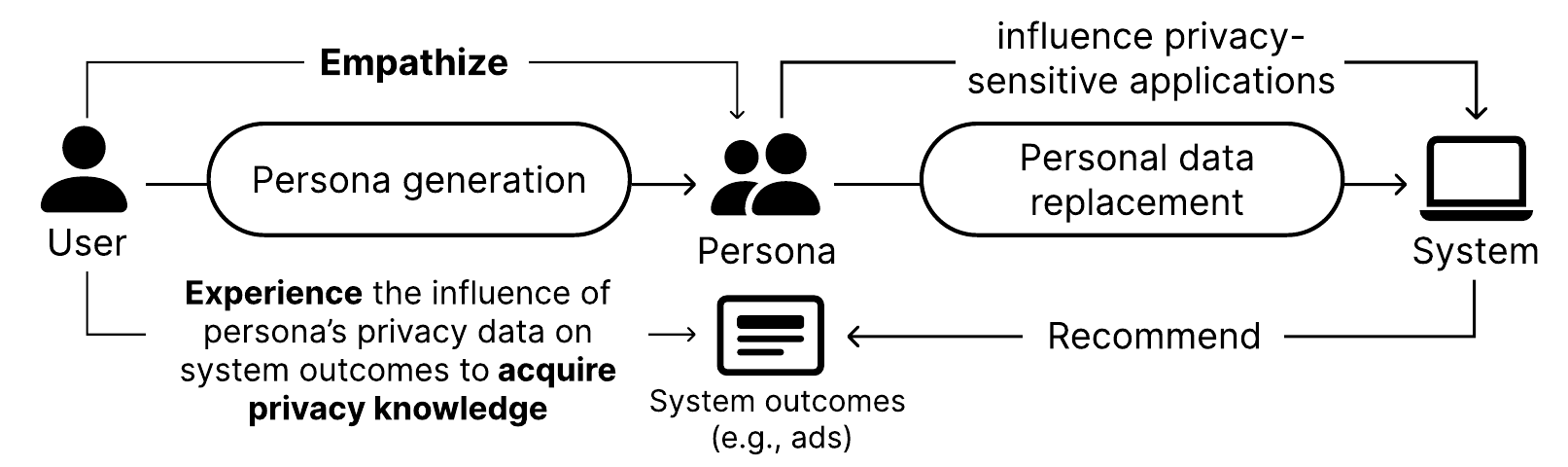}
    \caption{An empathy-based approach where users interact with online services by using the identity of different personas in a risk-free sandbox without leaking their real personal data. Users can cognitively and emotionally empathize with personas, observe and experience the causal effect between the privacy data and system outcomes (e.g., target ads), and acquire privacy knowledge.}
    \label{fig: approach}
\end{figure*}

Our approach has two core phases: persona generation and personal data replacement. Persona generation involves constructing personas with sufficient detail. Once completed, in personal data replacement, users will use personas to explore various privacy settings and online services from the perspective of a particular persona. An overview of this methodology can be seen in Fig. \ref{fig: approach}. 
\cc{Our focus on online advertisements as an example domain of system outcomes is motivated by their broad impact on everyday life, privacy sensitivity, and potential to raise privacy concerns \cite{zeng2021makes}. Targeted ads, a major component of online ads, usually track the user's online behaviors and personal information, but users often lack awareness of the specific privacy data that these ads automatically collect \cite{ur2012smart,kang2015my}. Note that online ads are just one possible manifestation of system outcomes. Other possible system outcomes include social media feeds, personalized news recommendations, and algorithmic decision-making, as detailed in Section~\ref{sec:downstream tasks}. We anticipate that our approach can be applicable to a wider variety of different system outcomes.}
In the following sections, we explain the two steps in detail, followed by introducing a prototype that integrates both steps---the Privacy Sandbox.

% \tlcomment{provide a few examples of other kinds of system outcomes here, expand the discussion in 6.6.4, and refer to 6.6.4 here.}

% \tlcomment{ads is only one of the possible types of system outcome here -- and we only ``chose'' ads for the study in Sec. 4. The entire Sec. 3 should talk about a broad range of possible domains.}

% \tlcomment{need to discuss the parts of how our approach leads to emotional, knowledge, and behavioral change at a high level}

% \yaxing{persona generation is more related to the empathy-based approach. We can talk about that first. Start with something like "Use personal to build empathy" where you explain why using persona, how persona can be used to build empathy, followed by a section called something like "from empathy to privacy literacy", and explain how empathy can help increase privacy literacy. Then, you talk about our implementation - to make the empathy-based approach work, we built a sandbox}

\subsection{Persona generation}
\label{sec: persona_generation}
The goal of the persona generation stage is to create artificial personas that contain realistic synthetic personal data.  By doing this, external applications and web services will read the synthetic data of the personas instead of the real data of actual users when they use the interactive sandbox. Consequently, these applications and services will tailor content based on the generated persona, letting users see the results of different privacy behaviors without risking their actual data.

% In the first stage, without risking leaking users' own personal information, our approach generates realistic personas that contain synthesized personally identifiable information, demographic information, longitudinal personal data, and other personal information. 

To effectively influence system outcomes and invoke user empathy, we chose specific data attributes to include in our personas.

% Below are the prompts we used to generate personas. Due to space constraints, we only display the most essential prompt here. The complete prompt can be found in \hl{Appendix X}.
% \tlcomment{move all prompts to the appendix -- maybe at most keep one to illustrate the format and the few-shot approach}
\subsubsection{Selected data attributes}
\label{sec:priacy_data_attributes}

\begin{itemize}
    \item \textit{Personally identifiable information (PII):} first name, last name, profile picture, and date of birth.
\end{itemize}

% Explain the rationale, why selecting each data
Our rationale for choosing these attributes is as follows.
Although recommendation systems are often based on anonymized data \cite{resnick1997recommender} and do not rely on profile pictures, names and profile pictures are still fundamental in personas \cite{grudin2002personas}. They make personas real and relatable, serving as vital stimuli of user empathy \cite{dziobek2008dissociation}. Furthermore, birth dates not only help establish the persona's age, making them more recognizable to a certain age group, but are also pivotal for personalized content \cite{miaskiewicz2011personas}. However, due to ethical concerns, we omitted sensitive PII like phone numbers and Social Security Numbers.

% \tlcomment{talk about how those attribute also contribute to the generation of longitudinal personal data}
% added

\begin{itemize}

    \item \textit{Demographic information:} age, gender, race and ethnicity, languages, education, income, occupation, home address, marital status, and parental status.

\end{itemize}

Demographic details, hobbies, and online interests play a crucial role in creating realistic personas. These attributes allow users to quickly connect with and relate to personas through shared characteristics \cite{miaskiewicz2011personas}. Such connections foster user engagement and empathy, making interactions with personas more meaningful and relatable. Furthermore, online recommendation systems often utilize demographic data and personal preferences to tailor their offerings \cite{mohamed2019recommender}. This kind of personal data guides how online platforms categorize users and, subsequently, the type of content they receive. When users observe how demographic information and personal preferences impact the services or user experiences, they are more likely to disentangle the system's opaqueness, understand how the system might use their privacy information, and consequently enhance their privacy awareness and literacy. Additionally, these data contribute significantly to the generation of longitudinal personal data. They influence a user's weekly schedule by reflecting their lifestyle choices and priorities. These data also shape one's browsing history and social media posts, as individuals tend to browse and share content related to their hobbies and demographic identities. 

\begin{itemize}
    \item \textit{Additional personal information:} devices in use, browser in use, hobbies, and online interests.
\end{itemize}

\cc{We incorporated details about devices, browsers, hobbies, and online interests into our personas, given their influence on online service personalization.} Notably, certain studies, such as the one by Nikiforakis et al. \cite{nikiforakis2013cookieless}, found that advertisers can use browser and device fingerprinting to tailor the ads for users. Hannak et al. \cite{hannak2014measuring} observed that e-commerce platforms might offer different prices depending on the device or browser of the user. \cc{Furthermore, user hobbies and interests are often gathered through methods such as website tracking, online surveys, and social media monitoring, aiding in customer segmentation}\cite{sari2016review}. \cc{Explicitly including such information in the persona allows the model to generate coherent person data (e.g., location history, browsing/search history, and social media posts) that reflect them. Therefore, when the user recognizes the system outcomes that relate to the hobbies and interests of the persona when interacting with the sandbox, they can understand their connection to the system's access of personal data, contributing to the privacy knowledge of the user.}

\begin{itemize}

    \item \textit{Longitudinal personal data:} weekly schedule with location logs, browsing history, and social media posts.
\end{itemize}

Unlike traditional UX personas, we incorporate longitudinal personal data, such as weekly schedules, location logs, browsing histories, and social media activities. From the user's perspective, this richer dataset paints a more comprehensive picture of the persona's life, fostering deeper user empathy. Previous work \cite{grudin2002personas} has also underscored the significance of longitudinal personal data in understanding personas. Rijn et al.'s study \cite{van2011achieving} emphasizes the value of in-depth behavioral data in understanding personas. Thus, by observing these personas' daily activities and interests, users can better empathize with them, seeing them as dynamic individuals with changing preferences.

From the system's perspective, these temporal data are pivotal for online services to make contextual recommendations. Several studies have illustrated the use of social media \cite{camacho2018social} and browsing histories \cite{tarus2017hybrid, beigi2019protecting, rajendran2021using} in predicting user preferences. Consequently, this data not only amplifies user empathy, but also impacts the tailored content they encounter. \cc{Such insights help users infer the connection between their data and the content they receive, contributing to empathy and privacy knowledge}

% We decided to generate such data to represent a persona for three reasons: (1) From the user's perspective, it aligns with the personas in user experience design, which emphasizes the necessity of personal and demographic information to foster users' understanding and emotional connection \cite{grudin2002personas, miaskiewicz2011personas}. (2) From a system perspective, many online service providers leverage similar personal information and longitudinal data to enhance recommendation systems and tailor services to user preferences \cite{narayanan2010myths}. (3) For highly sensitive data such as social security numbers, ethical and security protocols associated with large language models preclude their generation.

\subsubsection{Data generation methodology}

% \yaxing{I think we need a short paragraph here to explain that we used GPT-4 and prompt to generate the data. Otherwise there is a gap here. Or, simply move 3.2.3 here. Put a pointer to the appendix. }

% \tlcomment{I propose we can show (1) a diagram for the generation of the portrait image; and (2) one example prompt of social media posts to show the chain-of-thought and the few-shot methods. All other prompts go to the appendix}

To create comprehensive persona data, we introduce a novel pipeline that augments the outputs of large language models (LLMs) using few-shot learning~\cite{brown2020language}, contextualization~\cite{brown2020language}, and chain-of-thoughts techniques~\cite{wei2022chain}. 
\cc{Few-shot learning enables LLMs to quickly adapt to privacy persona generation using only a few examples. Contextualization allows LLMs to integrate additional context, such as weekly schedules, for more relevant output in tasks such as generating browsing history and social media posts. The chain-of-thoughts technique deconstructs complex tasks into simpler steps, enhancing the models' ability to manage intricate privacy data generation. These three techniques collectively enhance LLMs' performance by making them more adaptable, context-aware, and capable of generating complex data.}
For readers' reference, we have included detailed prompts and examples of few-shot learning in Appendix \ref{appendix:prompts}.

\textbf{Persona description:} The foundational step in our process is generating a personal description, which informs subsequent data generation to ensure alignment. We use a template prompt coupled with few-shot learning~\cite{brown2020language}, to guide GPT-4 in producing personally identifiable and demographic information. Users can customize the generation of their desired persona by providing other guidance as input.

\textbf{Privacy attributes:} 
We use GPT-4 and few-shot learning to parse the persona description generated and obtain attributes for each PII and demographic information to allow further modifications.

% \begin{figure}
%     \centering
%     \includegraphics[width=0.8\linewidth]{figures/draft_profile.png}
%     \caption{The generation pipeline of profile portrait images.}
%     \label{fig:profile}
% \end{figure}

\begin{figure*}
    \centering
    \includegraphics[width=1.0\linewidth]{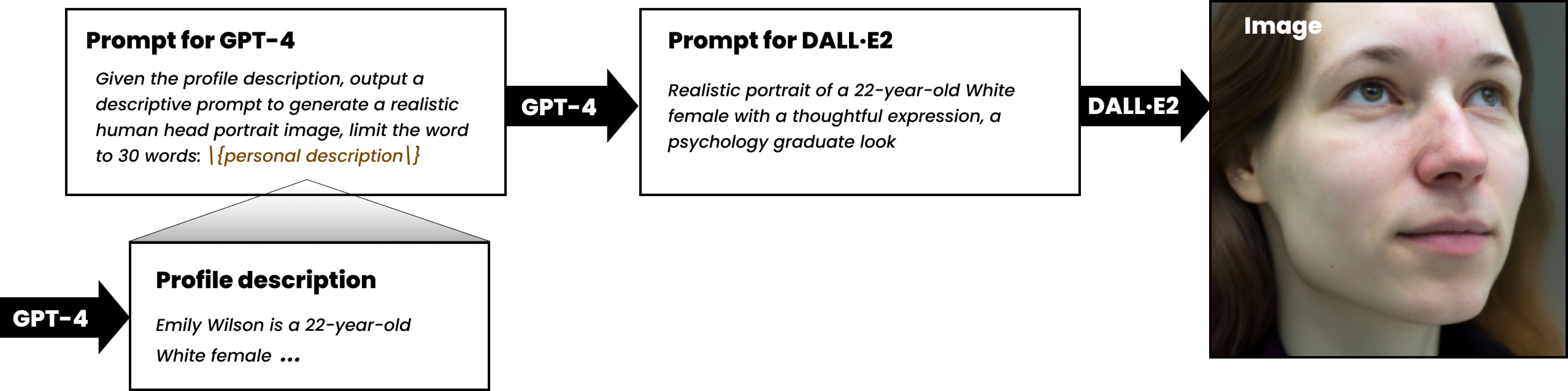}
    \caption{The generation pipeline of profile portrait images.}
    \label{fig:profile}
\end{figure*}

\textbf{Profile portrait image:} To make the generated persona feel more tangible and authentic to users, we employ a ``chain-of-thought'' approach \cite{wei2022chain} to create prompts for the generation of profile portrait images. As shown in Fig.~\ref{fig:profile}, we start by entering the personal description to generate a prompt for the OpenAI DALL·E 2 image generation API \footnote{https://platform.openai.com/docs/guides/images/image-generation-beta}. After obtaining the prompt, we then invoke the image generation API to synthesize a personal portrait image.

\textbf{Device and browser:}
Since device and browser information is typically contained within the browser's user agent, and user agent information does not appear directly in the personal description, we have separately created a prompt for GPT-4 to predict user agent, device, and browser based on the persona's personal description.

\textbf{Weekly schedule with location records:} 
A persona's weekly schedule provides insights into daily routines, further informing the creation of browsing histories and social media posts. By employing few-shot learning and contextualization, the persona's description is embedded within the prompt, ensuring schedule consistency. For geographical context, we have incorporated sample addresses into the few-shot learning examples, ensuring that generated events include reasonable location information.

\textbf{Browsing history:}
To generate consistent browsing history that aligns with the persona's personal description and weekly schedule with location records, we include them as the context in the prompt for generating browsing history for a specific time period. During the generation process, we also utilize few-shot learning to provide sample references for the browsing history records.

\textbf{Social media posts:}
To ensure realistic and consistent social media posts, we use the persona's description and weekly schedule as contextual anchors in the prompt. Typically, social media posts might contain visuals, so we randomly add 0-2 images per post. If an image is integrated, we employ the ``chain-of-thought'' technique similar to how we generate profile pictures: the content of the post serves as input, generating a prompt for image synthesis. This prompt is then fed to the OpenAI DALL-E image generation API. This process enriches the realism of the persona, aligning visual content with the textual content of the post. An illustrative prompt for social media posts is presented below.

% To ensure that social media posts align with the persona's personal description and weekly schedule with location records, we employ few-shot learning and use them as context in the prompt to generate post content. Since social media posts often include images, we randomly generate 0-2 images for each post. If a post includes images, we use the ``chain-of-thought'' method, inputting the generated post content to obtain a prompt for generating post images. We then call the OpenAI DALL·E image generation API to obtain the images. This approach enhances the realness of the persona by making the images consistent with the post content. We show an example prompt ofsocial media posts as follows.

\promptbox{
\textbf{Prompt for generating social media posts}
\vspace{5pt}

Provide ideas for this person to write posts (limit
the word to 140 words) based on the profile and location history: \{\textcolor{brown}{profile}\} \{\textcolor{brown}{location\ history}\}\\
Return a list of lists: \textcolor{blue}{<few-shot example posts>}\\
Output the posts in the following JSON format in plain text:
\{
    ``time'': <time in string format>,\
    ``address'': <address where this person shares the life>,\
    ``content'': <content>,\
    ``latitude'': <fake latitude>,\
    ``longitude'': <fake longitude>,\
    ``timezone'': <time zone>,\
    ``locale'': <locale>\
\}
}

\promptbox{
\textbf{Few-shot learning example for generating social media posts}
\vspace{5pt}

\textcolor{blue}{``posts'':
``[
    [``2023-06-01 08:31:10'', ``Starting my day with a delicious cup of coffee at my favorite coffee shop. Ready to conquer the world!  \#CoffeeLover \#MorningMotivation'', ``Coffee Shop - 123 Main Street, Brooklyn, New York 11207''],
    [``2023-06-01 18:00:34'', ``Just got back from the grocery store. Stocked up on essentials for the week. \#GroceryHaul \#MealPrep'', ``Grocery Store - 456 Broadway Avenue, Brooklyn, New York 11207'']
]''}
}

\promptbox{
\textbf{Prompt for generating a prompt for generating an image associated with the post}
\vspace{5pt}

Given the post \{\textcolor{brown}{content}\}, output a descriptive prompt to generate a realistic life image, limit
the word to 30 words:
}

\subsubsection{Implementation details}
To generate synthetic data, we used a Python script that interacted with the GPT-4 public API. We specified a maximum continuation length of $4,500$ tokens. Our approach to achieving few-shot learning involved utilizing the ``FewShotPromptTemplate'' available in the open-source Python library called ``langchain\footnote{ https://pypi.org/project/langchain/}''. Furthermore, we configured the GPT-4 model with a temperature parameter of $0.9$. The resulting images, generated using OpenAI DALL-E, were set to a size of $256\times256$ pixels.

% \tlcomment{don't call it identity modification -- maybe say personal data replacement? or personal data disguise if we want to emphasize the ``disguise'' aspect}

\subsection{Personal data replacement}
\label{sec:data_replacement}
In the second stage, users can use the identities of the generated personas to interact with various online services. The sandbox replaces users' demographic data within the Google account, real-time location, IP address, and web browsing history to match the persona's attributes. When an online service requests this personal information, the system offers the synthetic data of the persona. For the service provider, these data seem genuine, allowing them to provide personalized content as if they were interacting with a real user. This approach offers users a risk-free platform to cognitively understand the tangible consequences of their privacy choices and emotionally empathize with the persona in a convincingly interactive environment without exposing their actual personal data. The subsequent sections detail the process of personal data replacement. 

\textbf{Google account:} 
Since we choose online ads to represent system outcomes \cc{as explained in Section~\ref{sec:approach_overview}}, data replacement for Google accounts primarily pertains to information within the Google Ad Center. Google Ad Center's control portal\footnote{https://myadcenter.google.com/controls} allows users to customize the information provided to Google Ads, encompassing details of age, gender, language, relationship status, household income, education, industry, and homeownership. To substitute the profile data in the Google Ad Center with the privacy data of the persona, we create a Google account dedicated to the application. We use three open-source node.js libraries (``Puppeteer\footnote{https://pptr.dev/}'', ``Puppeteer-extra'', and ``Puppeteer-extra-plugin-stealth\footnote{https://github.com/berstend/puppeteer-extra}'') to automate the replacement of profile data. The replacement process consists of three steps: (1) After entering the personal profile page in Google Ad Center, we traverse the ``aria-labels'' of all elements of the page to identify the attributes that need to be replaced. (2) We extract the persona attributes from the database and process the data. (3) Then, we replace the values of target attributes with the persona's corresponding information.

% The control page in Google ad center\footnote{https://myadcenter.google.com/controls} allows the customization of the information provided to Google Ads, including age, gender, language, relationship status, household income, education, industry, and homeownership. To substitute the profile data in the Google ad center with the privacy data of the generated persona, we create a Google account dedicated to the application. We use three open-source node.js libraries (``Puppeteer''\footnote{https://pptr.dev/}, ``Puppeteer-extra'', and ``Puppeteer-extra-plugin-stealth''\footnote{https://github.com/berstend/puppeteer-extra}) to automate the replacement of profile data. The replacement process consists of three steps: (1) After entering the personal profile page in Google ad center, we traverse the ``aria-labels'' of all elements of the page to identify the attributes that need to be replaced. (2) We extract the persona attributes from the database and process the data. Since many attributes in Google ad center are categorical data, we use conditional logic to map the generated persona's data to the attribute categories. (3) Then, we replace the values of target attributes with persona's corresponding information.

\textbf{Geographical location:} Personalized online advertisements are often tailored based on the user's geographical location. Chrome browser supports location override. The replacement of the geographical location involves two steps. First, based on the persona's home address, we use the open-source OpenStreetMap's geocoding API\footnote{https://nominatim.openstreetmap.org/ui/search.html} to obtain the latitude and longitude of the generated persona's current address (based on their generated schedule). Then, we use the ``setGeolocation'' method from Puppeteer to modify the geographical location of the webpages based on the obtained latitude and longitude coordinates.

\textbf{IP address:}
We use NordVPN's API\footnote{https://nord.readthedocs.io/en/latest/reference/api.html} to modify the user's IP address. Based on the latitude and longitude of the persona's current location we obtained from OpenStreetMap, we calculate the nearest NordVPN server station to that location and select the server with the lowest load for connection. Once this connection is made, any online service that inquires about the IP address will receive the server's IP address instead of the user's original one.  % If none of the servers in the chosen station can be connected, we select the server from the station that is the second nearest to the persona's location.

\textbf{User agent:} 
To adjust the user agent, we employ the ``setUserAgent'' function in Puppeteer. This replaces the current page's user agent with the device and browser details associated with the generated persona.

\textbf{Browsing history:}
Chrome keeps its browsing history on the local computer using the SQLite database. Before launching the browser, we utilize a JavaScript script to overwrite the corresponding database file. Specifically, we substitute both the URL table, which logs visited links, and the visit table, which notes browsing timestamps, with the browsing history of the generated persona.

\subsection{A prototype for study: Privacy Sandbox}
\subsubsection{Privacy sandbox in action}
\begin{figure*}[!htbp]
    \centering
    \includegraphics[width=\linewidth]{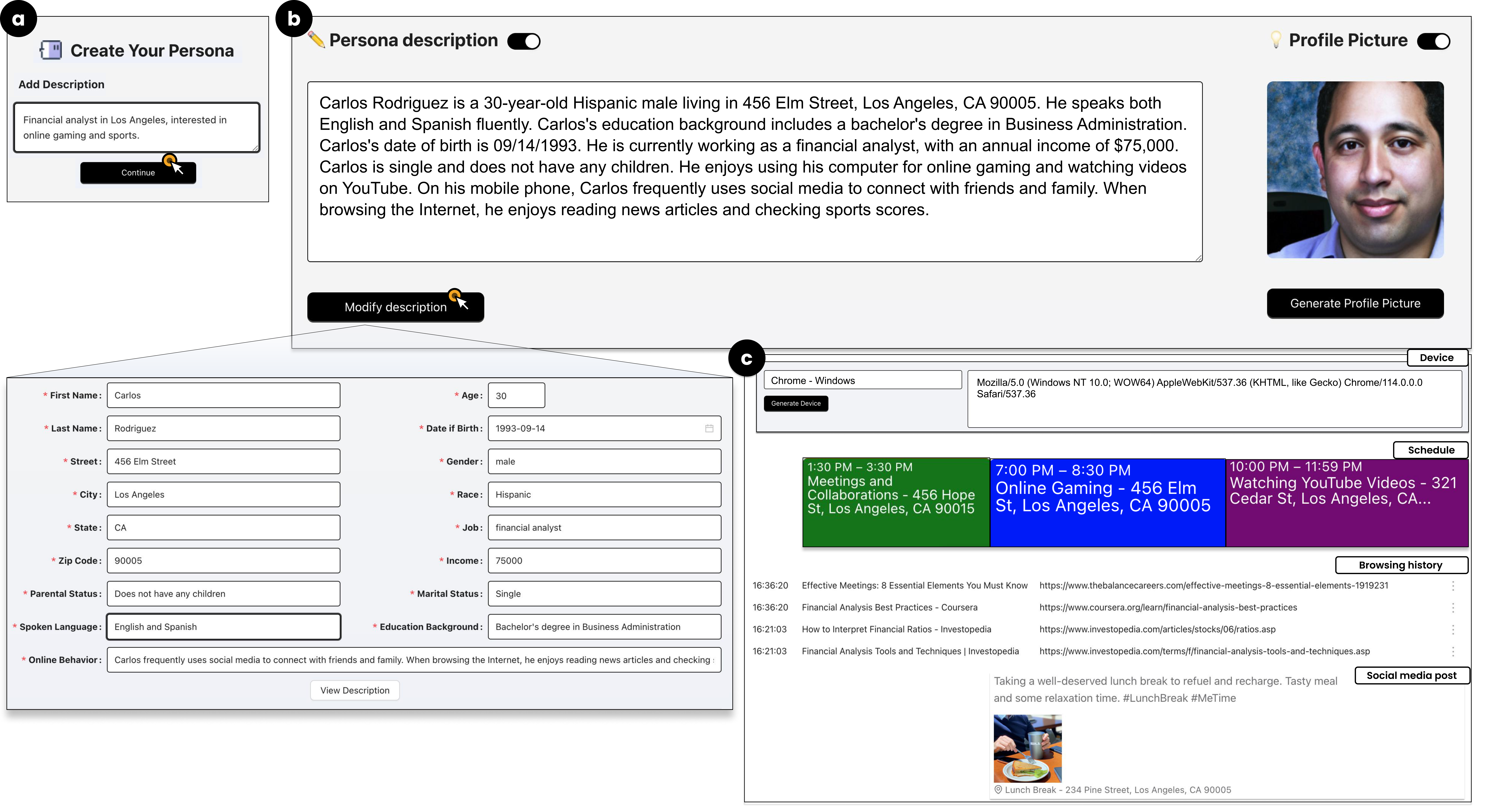}
    \caption{Privacy Sandbox User Journey. (a) Providing guidance for Persona's Profile Generation: The User's initial input acts as a seed for persona creation, exemplified by Bob's specific professional and personal interests. (b) Initial Persona Profile Generation and Customization: Creation of a preliminary persona ``Carlos Rodriguez'', which users can review and modify. (c) Generating additional privacy data aligned with the profile: Extension of the persona's attributes, ensuring alignment with the initial profile.}
    \label{fig: Scenario}
\end{figure*}

We demonstrate the use of the Privacy Sandbox through an example usage scenario.  In this scenario, a user creates a persona to navigate online services, showcasing the core features of Privacy Sandbox.  

Consider Bob, a financial analyst who wants to understand how private data impacts online ads. Using the Privacy Sandbox, he can generate a persona and act as the persona to browse websites that contain ads. 

\begin{enumerate}
\item \textit{Providing Guidance for Persona Profile Generation:}
Bob chooses the ``create a new persona'' button and enters his guidance to generate a persona that is similar to his profile. He enters ``Financial analyst in Los Angeles, interested in online gaming and sports.''
The ``guidance'' in this context acts as a seed or initial information. Users can provide as little or as much information as they feel comfortable with, ensuring flexibility while guarding their own private information. This information is not restricted to job titles or locations but could include hobbies, interests, or any other relevant information.

\item \textit{Initial Persona Profile Generation and Customization:}
Upon receiving the ``guidance'', the Privacy Sandbox generates a preliminary persona for Bob. The persona generated for Bob is named Carlos Rodriguez, a 30-year-old Hispanic male living in Los Angeles. Carlos speaks both English and Spanish and has a bachelor's degree in Business Administration. He works as a financial analyst and earns an annual income of \$75,000. He enjoys online gaming, watching YouTube videos, and checking sports scores.
At this point, Bob can review the generated persona and modify any attributes of the persona, as shown in Fig. \ref{fig: Scenario} (b).

\item \textit{Generating Further Privacy Data Aligned with the Profile:}
After Bob is satisfied with the profile, he proceeds to generate the detailed attributes of the persona. This includes the persona's device and browser in use, weekly schedule with location records, browsing history, and social media posts. Each part is generated to be consistent with the persona's profile. Bob has the option to modify or regenerate any part of these attributes.

\begin{figure*}[!htbp]
    \centering
    \includegraphics[width=\linewidth]{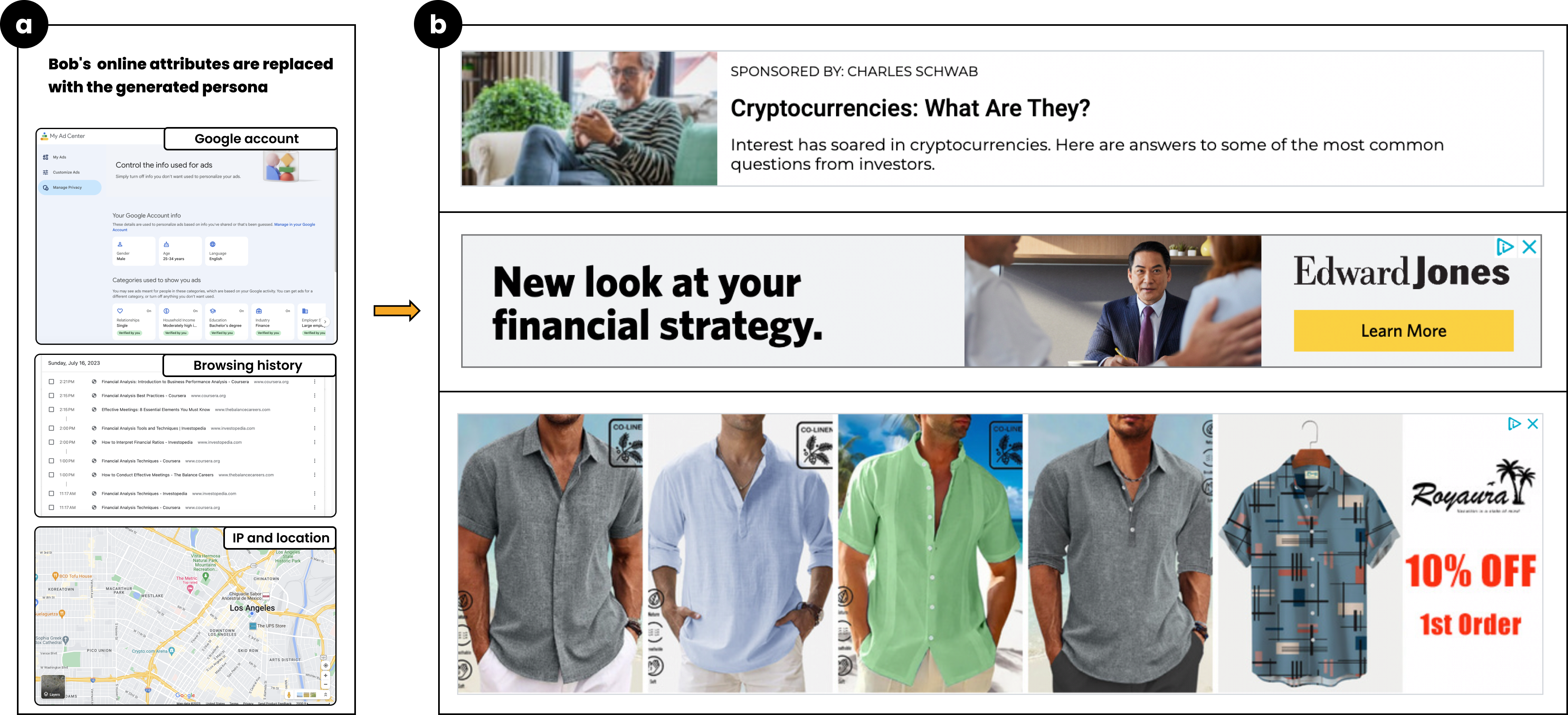}
    \caption{Browsing Online Services with the Generated Persona: Users, \cc{after activating the Chrome extension using their generated persona,} can interact with online services, observing the persona's influence on targeted ads.}
    \label{fig:Scenario}
\end{figure*}

\item  \textit{Browsing Online Services with the Generated Persona:}
Once Bob is satisfied with the persona, he can save this persona for future use. \cc{When he clicks on the ``activate'' button, the Privacy Sandbox will activate a Chrome extension that replaces Bob's privacy data with the synthesized data of Carlos, the artificial persona.} Bob's privacy data, including his profile in the Google Ad Center, browsing history, real-time location, and IP address, are temporarily replaced. Bob can interact with websites and online services as usual, but online service providers will see him as Carlos and start providing him with personalized ads, customized content, and algorithmic decisions they would give to a person like Carlos. Bob can then experiment with different privacy settings (e.g., enabling/disabling the access to certain data for a website) and behaviors (e.g., visiting certain sites when the visits are tracked, voluntarily providing personal data to a service), experiencing how his user experience has changed as a result. For example, he may start to observe seeing ads customized based on the attributes of Carlos.
\end{enumerate}

\subsubsection{Implementation}
We developed the Privacy Sandbox with a React-based fronte-nd and a back-end powered by Flask and SQLite3. They communicate through HTTP requests for API access. The SQLite database stores different types of synthesized personal privacy data: persona profiles, schedules, browsing history, Twitter posts, and Facebook posts. 

We commit to open-sourcing our implementation of the Privacy Sandbox, the Chrome plugin for browser data replacement, as well as the persona data generation pipeline.

\section{User Study}
We conducted a user study\footnote{The study protocol was reviewed and approved by the IRB at our institution.} with 15 participants to evaluate our approach. The study examined the following research questions:
\begin{itemize}
    \item \textbf{RQ1:} How realistic are the artificial privacy personas generated using our approach, in comparison to real personal data and the baseline GPT-generated data?
    \item \textbf{RQ2:} How do the different characteristics of the synthesized privacy data of personas impact the user-perceived realness?
    \item \textbf{RQ3:} Can our approach of replacing user personal data with our synthesized data of personas invoke changes in system outcomes?
    \item \textbf{RQ4:} Can users invoke empathy and perceive the links between privacy data and system outcomes when using the Privacy Sandbox?
\end{itemize} 

\subsection{Participants}
We recruited 15 participants through word-of-mouth, LinkedIn, and Twitter. Eight of them participated in the study in-person at a usability lab and seven participated virtually through Zoom. Participants were required to complete a pre-screening survey to collect their basic demographic information, including age, gender, state of residence, and race/ethnicity. We tried to diversify the participant group as much as possible. The demographic information of our 15 participants is shown in the Appendix \ref{appendix:participant_demographic}. Our participants' age ranges from 19 to 33, with nine females and six males. Each participant was compensated with \$40 USD for their time.

\subsection{Study design}
Each study session lasted around 90 minutes. The session consisted of three phases.

\subsubsection{Study procedure}
After the informed consent process and a brief introduction to the study, each participant went through the following three phases of the study procedure: 

\begin{itemize}
\item \textbf{Phase 1: Quantitative evaluation of generated personas} Participants were presented with three personas on our developed privacy sandbox platform, one each randomly chosen from three distinct groups: personas generated with our approach, real personas, and personas directly generated using the GPT-4 model \cc{(detailed in Section \ref{sec:persona_group}). Each group contains eight personas.} The order of the personas was randomized. Participants were tasked to rate each persona's clarity, completeness, credibility, consistency, and level of empathy using the five-point Likert scale. 

\item \textbf{Phase 2: Qualitative investigation of generated personas} To gain a deeper understanding of user perceptions of generated personas and how real participants perceive different parts of them, we adopted a combined ``Think Aloud"~\cite{jaaskelainen2010think} method and semi-structured interview approach. The experimenter first introduced the use of the privacy sandbox to the participants, ensuring that they understood how to generate and modify personas. Afterward, the selected generated personas were presented to the participants. As they navigated the personas, participants were instructed to vocalize their overall impressions and specifically comment on the elements of the persona's privacy data \cc{(as detailed in Section \ref{sec:priacy_data_attributes} and Figure \ref{fig: Scenario})} enhanced or diminished the sense of realness. During the ``Think Aloud" process, as participants shared their immediate feedback, researchers could interject with follow-up questions or ask for clarifications. If participants identified certain elements as inauthentic or felt adjustments were needed, the interface allowed them to directly modify the persona profiles, including attributes, avatars, weekly schedules, browsing history, and posts. They could either make direct modifications to the interface or verbally describe the desired changes. For every modification or suggestion, participants were asked to explain their reasoning. Each participant was exposed to two personas, counterbalanced, and selected from the eight personas. A detailed list of all the important attributes of these eight personas can be found in the Appendix \ref{appendix:our_personas}.

\item \textbf {Phase 3: Analyzing ad-persona connections} In this phase, participants completed the task of correlating the persona information with the advertisements on given websites. The goal of this phase is to investigate whether users can perceive the correlation between privacy data and system outcome (e.g., target ads). Each participant completes the task for two personas that they have not seen before, randomly selected from the personas generated by our mechanism. For each persona, first, the participant read a persona using the privacy sandbox prototype. After reading, they clicked on a designated ``active'' button. This triggered the launch of a new browser window by the sandbox that automatically replaces their real personal data with the persona's synthetic data. As explained in Section \ref{sec:data_replacement}, the sandbox replaced persona attributes, browsing history, location, and IP address (as seen in Figure \ref{fig:Scenario}). Then, the participant was asked to read the home pages of two websites for each persona, randomly chosen from the five websites shown in Table \ref{tab:website_list}. Participants were tasked to identify ads that are targeted to the current persona, record them in a spreadsheet, and explain how the ads relate to the persona in a think-aloud manner. 
\end{itemize}

\subsubsection{Personas}
\label{sec:persona_group}
We prepared three groups of personas for the study: (1) artificial personas generated using our approach; (2) real personas collected from users; and (3) personas generated directly using the GPT-4 model. Each group contains eight personas. 

\textbf{Personas generated with our approach:} Using our proposed approach (described in Section \ref{sec: persona_generation}), we generated personas using a diverse range of demographic attributes such as age, city, educational background, and gender (See Appendix \ref{appendix:our_personas} for the full details).
    
\textbf{Real personas:} We recruited eight adult participants to create a sample set through word-of-mouth and social media including LinkedIn and Twitter. 
We collect the same list of information as the list of synthesized privacy attributes for artificially generated personas. All participants were fluent in English, had active Facebook and Twitter accounts, and were willing to share their posts and browsing data for the past week. The group had diverse demographics as shown in Table ~\ref{tab: real_persona}. 

\begin{table}[ht]
\small
\centering
\begin{tabular}{cccccc}
\toprule
ID & Age & Gender & Education level & Ethnicity & Digital Literacy\\
\midrule
1 & 27 & Female & Master               & Asian                  & 5\\
2 & 19 & Female & Bachelor             & Asian                  & 5\\
3 & 20 & Male   & High school   & White        & 5\\
4 & 28 & Female & Master               & Asian                  & 4\\
5 & 25 & Male   & Bachelor             & Black & 5\\
6 & 28 & Female & Master               & White        & 5\\
7 & 23 & Female & Master               & Hispanic        & 4\\
8 & 33 & Male   & Ph.D.                & Asian                  & 5\\
\bottomrule
\end{tabular}
\caption{The demographic information of real personas. The digital literacy was self rated on a 5 Likert scale where 1 stands for ``not at all proficient'' and 5 represents ``highly proficient''.}
\label{tab: real_persona}
\end{table}

To preserve their anonymity, we took the following measures to strike a balance between protecting their privacy and maintaining the perceived realness of the personas:
\begin{enumerate}
    \item We replaced any data disclosing their actual names with pseudonyms. When generating pseudonyms, we generated names that align with the cultural background of the persona based on their race/ethnicity.
    \item We used generated profile pictures based on their age and race/ethnicity to replace their real portrait.
    \item We replaced their real addresses with fictitious ones that plausibly resembled their actual locations.
    \item We examine the browsing history and social media posts collected to anonymize entries that contained sensitive personally identifiable information.
\end{enumerate}

\textbf{GPT-generated personas:} 
% For our GPT-based personas, we leveraged the capabilities of the GPT model to generate 8 diverse personas in line with our predetermined guidelines.
To compare the quality of our persona generation pipeline with that of using GPT-4 directly, we generated eight personas using GPT-4 without using the few-shot learning, contextualization, and chain-of-thoughts techniques proposed in this paper. We used similar input guidances to ensure a diverse representation of the generated results. We demonstrate the baseline prompts to generate social media post content for personas in this condition as an example. The complete prompts are provided in Appendix \ref{appendix:prompts}.

\promptbox{
\textbf{Prompt to generate social media post content for GPT-generated personas}
\vspace{5pt}

Provide ideas for this person to write posts (limit
the word to 140 words).\\
Output the posts in the following JSON format in plain text:
\{
    ``time'': <time in string format>,\
    ``address'': <address where this person shares the life>,\
    ``content'': <content>,\
    ``latitude'': <fake latitude>,\
    ``longitude'': <fake longitude>,\
    ``timezone'': <time zone>,\
    ``locale'': <locale>\
\}
}

\subsubsection{Websites}

We selected five representative websites (as shown in Table~\ref{tab:website_list}) to test the Privacy Sandbox with personalized advertisements. We adopted the method previously used by Zeng et al.~\cite{zeng2022factors} to curate the sample websites. Our selection criteria encompassed the following aspects: 1) inclusion of a diverse range of website topics, 2) presence of multiple advertisements on the chosen websites, and 3) advertisements are sourced from Google Ads. This choice was motivated by (1) Google Ads is by far the most popular advertisement platform on the Internet with the reach of over two million websites and apps and over 90\% of Internet users worldwide\footnote{\url{https://support.google.com/google-ads/answer/117120?hl=en}}; (2) Google Ads has comprehensive access to personal data stored in the Chrome browser (e.g., browsing history, Google accounts) used in our study.  

\begin{table}[ht]
\small
\centering
\begin{tabular}{llrr}
\toprule
Website & Topics & \# of Ads & Site Rank \\
\midrule
www.weather.com & Weather forecasts & 8 & 37 \\
www.cnn.com & National news & 8 & 89 \\
www.researchgate.net & Academic paper & 3 & 556 \\
www.usnews.com & National news & 3 & 1,165 \\
www.fashionista.com & Fashion & 5 & 78,490 \\
\bottomrule
\end{tabular}
\caption{Websites visited by participants in the study}
\label{tab:website_list}
\end{table}

\subsection{Data analysis methods}

To analyze the quantitative data collected from the Likert scale survey, we employed one-way ANOVA for each survey item to evaluate the significant difference in mean scores regarding the personas' realness among three groups: our approach, the baseline GPT, and real personas. Whenever significant differences emerged, we conducted post-hoc tests using Tukey's pairwise comparisons to gain deeper insights into these distinctions.

For qualitative data analysis, we followed established open coding procedures \cite{brod2009qualitative}. Two members of our research team independently initiated the coding process in MAXQDA. A researcher coded 20\% of the sample and generated a set of initial codes. Subsequently, the second researcher coded the same portion to introduce new codes if necessary. Non-agreement cases were discussed to reconcile differences and establish a cohesive codebook.
Utilizing this codebook, we conducted a thematic analysis to uncover and delineate the significant themes that emerged during the interviews and were relevant to the established codes. The complete codebook is presented in Appendix \ref{appendix:codebook}. These themes were then consolidated and evolved into study findings that are detailed in Section \ref{sec:results_findings}.

\section{Results and Findings}
\label{sec:results_findings}

\subsection{Users' perceived realness of privacy personas \cc{(RQ1)}}

\begin{figure*}[!htbp]
    \centering
    \includegraphics[width=\linewidth]{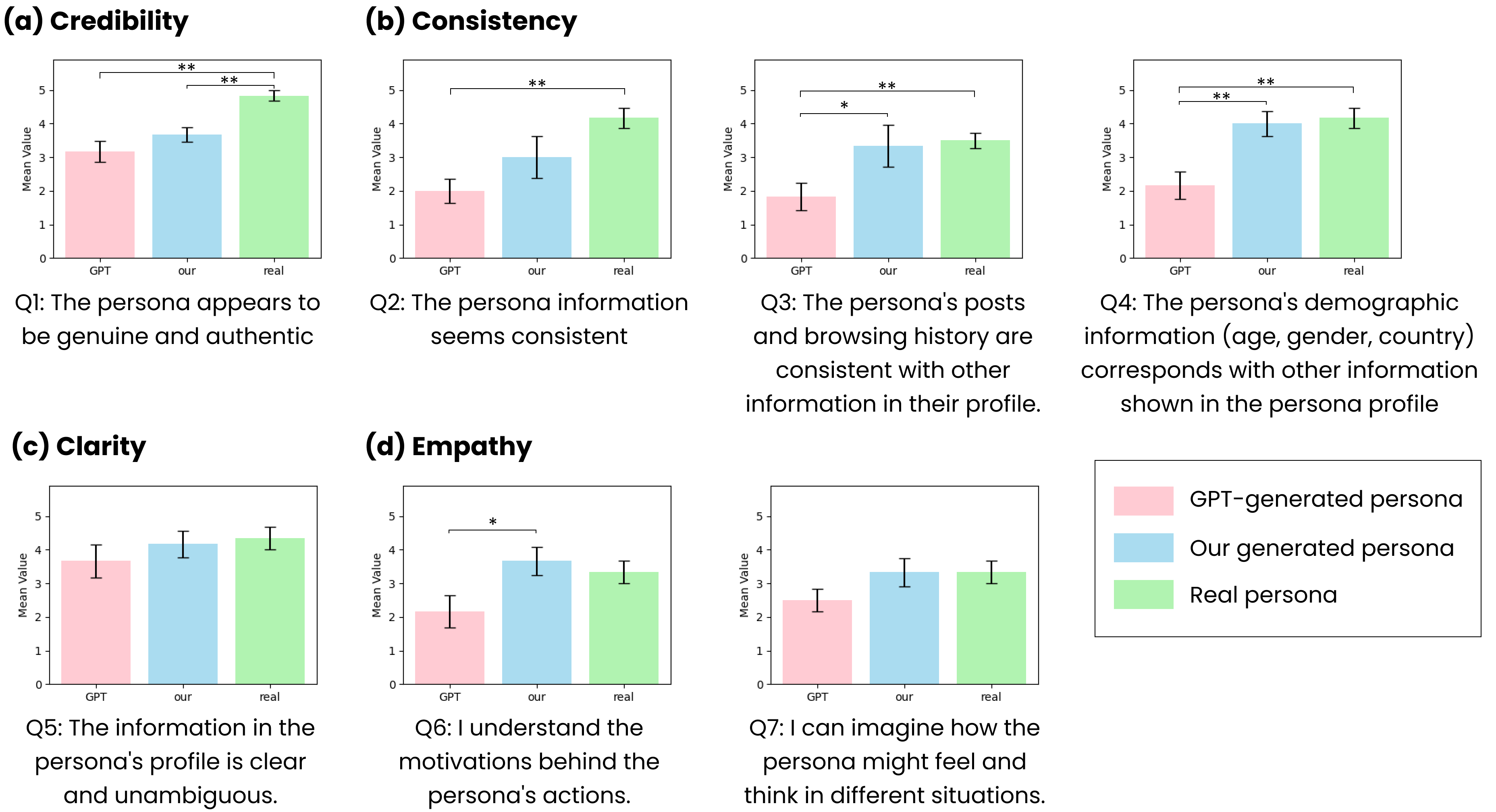}
    \caption{Means and standard errors of each measure in three conditions: GPT-generated persona, our generated persona, and real persona. All items are measured by user ratings on a 5-point Likert scale.}
    \label{fig:quant_data}
\end{figure*}

We applied one-way ANOVA and post-hoc tests to analyze the difference in perceived realness of users for three groups of personas. Fig. \ref{fig:quant_data} shows the mean and standard error for each measure, including credibility, consistency, clarity, and empathy. 

\subsubsection{Credibility:} 
Significant differences ($p<0.05$) in credibility were found between real personas ($M=4.83, SD=.408$) and GPT-generated personas ($M=3.17, SD=.753$) and between real personas and personas generated with our approach ($M=3.67, SD=.516$). No significant differences were observed between the GPT-generated personas and the personas generated with our approach. These results imply that while the persona generated by our approach received higher rating scores compared to the GPT-generated persona, the current generation models still fall short of achieving the level of credibility found in real personas.

\subsubsection{Consistency:} 
All measures indicate significant differences ($p<0.05$) in consistency between real personas (Q2: $M=4.17, SD=.753$, Q3: $M=3.50, SD=.548$, Q4: $M=4.17, SD=.753$) and GPT-generated personas (Q2: $M=2.00, SD=.753$, Q3: $M=1.83, SD=.548$, Q4: $M=2.17, SD=.983$), while there are no significant differences ($p<0.05$) in consistency between real personas and personas generated by our approach (Q2: $M=3.00, SD=1.549$, Q3: $M=3.33, SD=1.506$, Q4: $M=4.00, SD=.894$). Although there are no significant differences ($p<0.05$) between GPT-generated personas and personas generated by our approach in the overall consistency assessment (Q2), significant differences (Q3: $p<0.1$, Q4: $p<0.05$) are observed in the consistency of specific privacy attributes between them. 

\subsubsection{Clarity:} 
No significant differences ($p<0.05$) in information clarity are found between real personas ($M=4.33, SD=.816$), GPT-generated personas ($M=3.67, SD=1.211$) and personas generated with our approach ($M=4.17, SD=.983$). % These results imply that the personas generated by existing models can nearly match real individuals in terms of information clarity, as users can generally understand the meaning of persona privacy attributes when browsing through all three types of personas.

\subsubsection{Empathy:} 
A significant difference ($p<0.05$) in cognitive empathy (Q6) is found between GPT-generated personas ($M=2.17, SD=1.169$) and personas generated by our approach ($M=3.67, SD=1.033$). No significant differences are observed between real personas ($M=3.33, SD=.816$) and GPT-generated personas and between real personas and generated by our approach. No significant differences ($p<0.05$) in emotional empathy (Q7) in information clarity are observed among real personas ($M=2.50, SD=.837$), GPT-generated personas ($M=3.33, SD=.816$) and personas generated with our approach ($M=3.33, SD=1.033$).

Our results suggest that personas created using our method improve users' understanding of the personas' motivations when compared to those generated by GPT. However, the behavior of real personas is influenced by intricate factors. This complexity may cause users to exhibit slightly reduced cognitive empathy for real personas compared to those we generated. Interestingly, users showed no significant difference in emotional empathy across the three persona categories. The overall empathy scores (both cognitive and emotional) were moderately low. This could be because users only reviewed the profiles and did not immerse themselves in the personas' identities to experience the impact of privacy attributes on system outcomes. However, we observed that users expressed noticeable excitement or surprise when using personas' identities and encountering highly relevant advertisements (details in Section \ref{sec:perceived_links}). This suggests that relying solely on browsing personas' information has limited efficacy in eliciting empathy. The actual interactive experience of using personas through the Privacy Sandbox might be necessary to foster greater empathy towards personas.

\subsection{Factors influencing the perceived realness of generated personas \cc{(RQ2)}}
\subsubsection{Familiarity with the generated persona} \label{section: familiarity} 

Participants' perceptions of a persona's realness often correlated with their familiarity with that persona. For example, a participant from the financial sector, upon reviewing two personas (a financial analyst and a designer), remarked, ``\textit{I think this one (designer) is better than the last one (financial analyst)... Perhaps because I'm familiar with financial analysts.}'' Variability in familiarity with the same persona can lead to differing views on its realness. To illustrate, concerning the persona of a psychology research assistant, one participant felt the profile details matched the persona, saying, ``\textit{I think it (the schedule) is pretty much consistent with the personal information.}'' Yet, another participant expressed skepticism about the given work schedule, commenting, ``\textit{I think her working time is a little bit short. I expect (this) because I know some research assistants. I think they are busy.}''

% We observed that participants have higher expectations of realness in the privacy data of generated personas if they are more familiar with the personas. For instance, a participant working in the financial field, after browsing two personas (a financial analyst and a designer), mentioned, ``\textit{I think this one (designer) is better than the last one (financial analyst)... Perhaps because I'm familiar with financial analysts.}'' % (audio1202769299)
% The difference in familiarity with the same persona can also result in varying perceptions of its realness. For example, regarding the persona of a psychology research assistant, one participant commented, ``\textit{I think it (the schedule) is pretty much consistent with the personal information.}''  % (audio1987928654)
% However, another participant questioned the realness of the schedule, stating, ``\textit{I think her working time is a little bit short. I expect (this) because I know some research assistants. I think they are busy.}''  %(audio1659845317) 

\subsubsection{Deviation from personal experiences}
\label{sec:deviation_experiences}
Participants perceived personas as realistic when their behavior matched their own personal experiences, often resulting in positive feedback.  For example, one participant stated ``\textit{I think this day looks good because it looks like he was working this whole time until he went home and then watched YouTube and, um, like exercise.}'' %(audio1519478401). 
However, when there were discrepancies between the personas and the participant's own experiences, it led to skepticism or dissatisfaction. These discrepancies arise in two main contexts:

% \hl{1. Can I summarize them as stereotypes and common sense? 2. Should I explain how they establish such personal experience? like from hearsay, self-experience, or observing someone else?}

\paragraph{Discrepancies in Perception of Specific Individuals} 

This occurs when a persona does not match a participant's perception of a certain group of people. The discrepancies could relate to specific attributes or be more general. 

\begin{enumerate}
    \item \textit{Specific Attributes: } Participants questioned the realness of a persona if certain details did not align with their experiences. For instance, after seeing personas of a financial analyst and a cashier, six participants felt the given income was too low. A participant noted, ``\textit{One thing that I would notice is usually financial analysts make a lot of money. So \$70,000, this annual income does not seem reasonable. This seems unusual to me, given that Michael was born in 1981, he is 40 years old, so he probably has a lot of experience in the field. He definitely should have been making more than \$70,000.}'' Another issue raised was about browsing history; participants felt the content was too basic for an experienced individual, with one commenting, ``\textit{Given his age, I think he is an experienced custodian, does not need to search for the information about this job itself.}''

    \item \textit{Overall Impression: } At times, the deviation from personal experience was not attributed to a specific privacy attribute but rather to participants' overall impression of the persona. One participant, after reviewing all of a persona's information, said, ``\textit{This is like a fake person someone's trying to learn human behavior...It's like the whole thing looks too perfect to be real...It's like intentionally proving I'm doing this...Make me feel she's being controlled? Probably by her husband and just posing those things to show I'm alive}'' % (audio1705700564).
    This overall deviation often triggers users to question the realness of the persona more deeply.
\end{enumerate}

\paragraph{Discrepancies in Privacy Attributes} 
This pertains to inconsistencies in specific details when compared to the participants' own experiences. Discrepancies were observed in weekly schedules, browsing histories, and social media posts. 

\begin{enumerate}
    \item \textit{Weekly Schedule: } All participants felt that the work hours for the personas were too lengthy, with observations such as, ``\textit{It looks like they work at least 8 hours a day. Notice that I mean yeah 8 hours a day for seven days a week.}''

    \item \textit{Browsing History: } Four participants believed that browsing history was too centered on work, lacking diversity. One commented,  ``\textit{The website she is browsing is all about her job uh is all related to her job and her professional, but uh they are. There should be some other content about her life.}'' They also expected more continuity in browsing, mentioning, ``\textit{I would continue to click the content in those websites. So again, those links, those four links should be the same...should be consistent or should be progressive.}'' Participants were also skeptical of identical timestamps on different records, with a remark, ``\textit{She cannot be on the same page at the exact same time. This is to the second exact same time. This is incorrect.}''

    \item \textit{Social Media Posts: }   Some participants felt that the posts were overly positive and superficial. One observation was, ``\textit{I feel strange about his posts is that he always appears so positive, like his life is so perfect and he's very proud of his work.''} They thought that the contents were superficial, and lacked depth in emotions and thoughts, as one participant mentioned, ``\textit{There's no real emotion...it just feels like she wants to prove something to you.}''
\end{enumerate}

\subsubsection{Consistency within the data}
\label{sec:data_consistency}
Our generated persona is generally consistent across various privacy data attributes, a feature that participants frequently acknowledged and praised. For example, some remarked, ``\textit{The Facebook and Twitter posts are pretty consistent with the content in calendar timing.}'' and ``\textit{It seems very real and then she started browsing at 6:00 am.}'' However, certain inconsistencies in the generated data made some users question the realness of the persona. These inconsistencies can be broadly classified into two categories: direct data inconsistencies and out-of-context generated content.

\begin{enumerate}
    \item \textit{Direct Data Inconsistencies: } These inconsistencies can be observed within a single data type or across different attributes. For instance, many participants noted that in the weekly schedule, the same event occurred at different locations every day (e.g., ``\textit{his workout location changes every single day}''), or in social media posts, different images within the same post depicted inconsistent people or scenes (e.g., ``\textit{It's weird because for her desk changes across the pictures}''). % (audio1854517361) 
    Regarding inconsistencies across different attributes of data, participants found inconsistencies between the browsing history and the schedule (``\textit{In the afternoon and evening, he's not using the internet or his mobile phone. And in the schedule, it should have more history about liking YouTube videos, social media interaction}''). %(audio1632706610)
     This indicates a need to strike a better balance between randomness and diversity in content generation through large language models for future work.

     \item \textit{Out-of-Context Content: } Users identified certain generated browsing histories as not aligning with the persona's description. For instance, ``\cc{\textit{there are many histories about bike riding, but in his profile, there is no information about riding a bike.}}'' Such discrepancies were attributed to sample leakage during the few-shot learning process. For example, the model was exposed to browsing samples related to bike garages. To enhance the realness and relevance of generated data, we must address such technical issues in future iterations.

\end{enumerate}

\subsection{System outcomes influenced by privacy data replacement \cc{(RQ3)}}
\label{sec:system_sensitivity}
% two researchers
To assess the changes in system outcomes as a result of privacy data replacement, we calculated the advertisement overlap rate for eight generated personas across five selected websites using the following method: 
\[
\text{ad overlap rate} = \frac{\text{number of duplicated ads between personas}}{\text{total number of ads in a website}}
\]

The rationale behind the metric is that: if the privacy data replacement approach is effective, users should see distinct advertisements when they switch between different personas.

For every website considered, we began by gathering all the advertisements presented when accessing the site with each of the eight personas. Among these ads, we specifically noted those that appeared for multiple personas, effectively highlighting the number of ads that overlap or duplicate across personas.

\begin{table}[h]
\small
\centering
\begin{tabular}{lrrr}
\toprule
Website & duplicated ads & total ads & ad overlap rate \\
\midrule
www.weather.com & 22 & 47 & 46.81\% \\
www.cnn.com & 9 & 60 & 15.00\% \\
www.researchgate.net & 8 & 25 & 32.00\% \\
www.usnews.com & 8 & 21 & 38.10\% \\
www.fashionista.com & 16 & 36 & 44.44\%\\
\bottomrule
\end{tabular}
\caption{The ad overlap rate for each website.}
\label{tab: ovelap_rate}
\end{table}

\begin{figure*}[!htbp]
    \centering
    \includegraphics[width=1.0\linewidth]{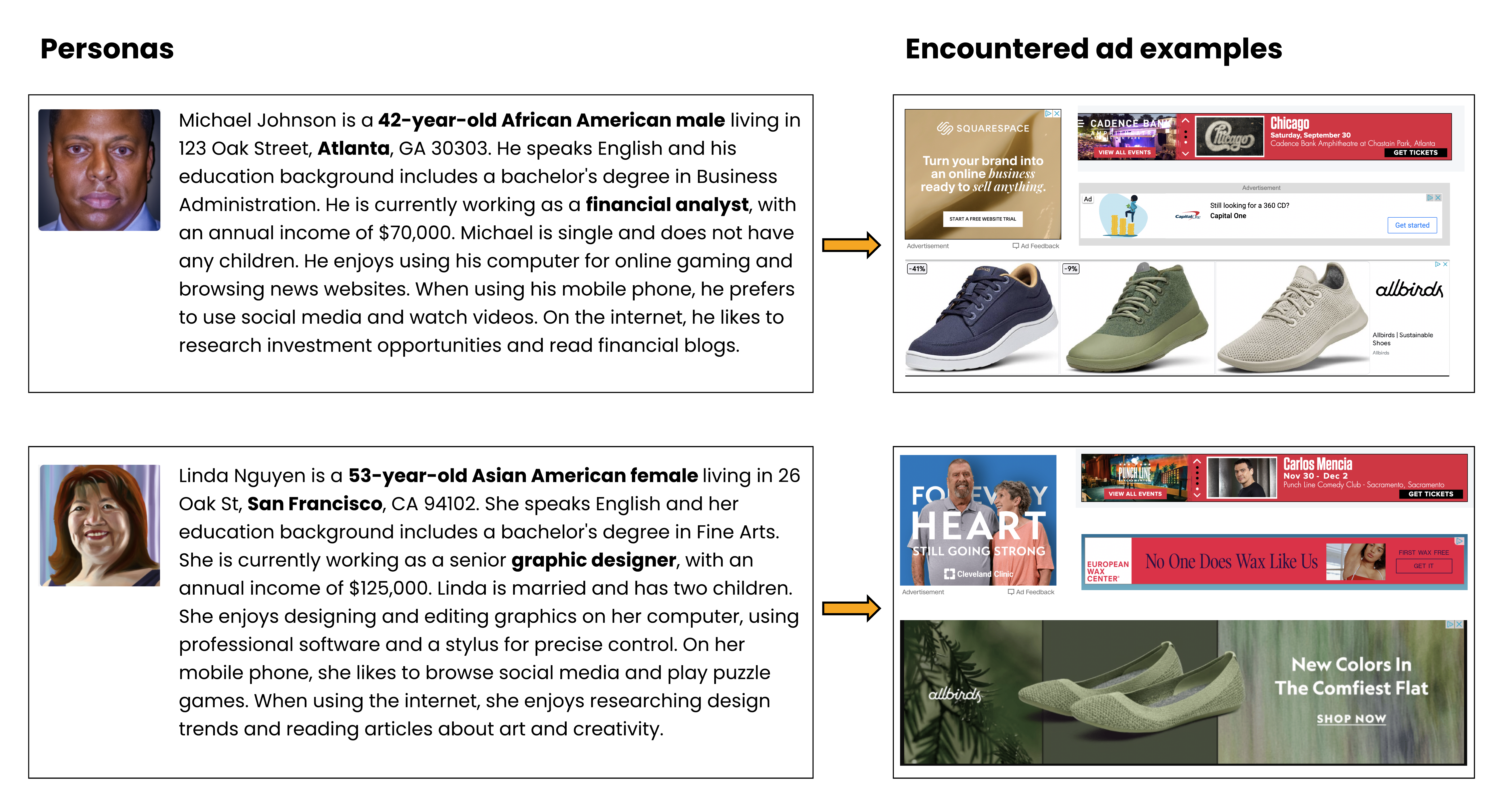}
    \caption{Examples of different ads encountered by participants when browsing the selected websites using various persona identities. Note that participants who used the first persona received ads about shows in Atlanta (associated with the location), investment tools (associated with the profession and browsing history), and shoes for men (associated with gender), while participants who used the second persona received different ads associated with her attributes.}
    \label{fig: persona_ad_examples}
\end{figure*}

% \tlcomment{for the overlap rate -- is this the avg. overlap rate across all users? If so, it should be clearly indicated in the description and formula}

Table \ref{tab: ovelap_rate} shows the result of the ad overlap rate for each selected website. The overlap rates for all websites are less than $50\%$. This implies that when users switch between eight personas, more than half of the ads they encounter are unique to a single persona. Fig. \ref{fig: persona_ad_examples} shows the variations in the ads on weather.com for different personas.  In subsequent sections, we will dive deeper into how users associate these advertisements with the underlying private information of the personas.

\subsection{Perceived links between privacy data and system outcomes \cc{(RQ4)}}
\label{sec:perceived_links}
All participants, while browsing the websites as personas, encountered ads related to the persona's privacy attributes. \cc{When inquired about the ads,} participants perceived the connections between the ads and the persona in two ways. 

\textit{(1) Direct connection based on privacy attributes:} Participants noticed an explicit correspondence between the ads and the persona's privacy attributes. Participants often associated ads with personas based on their interests, hobbies, daily activities, occupation, educational background, marital status, and family situation. For example, one participant claimed that an Xfinity ad was relevant to the selected persona because ``\textit{he does a lot of social media, gaming, YouTube. So he maybe wants to use Xfinity to watch something.}'' % (audio1931526300)
When an ad seemed especially pertinent to a persona, it evoked reactions of excitement or surprise.  For instance, when a participant was browsing the website through a persona living in Los Angeles and saw an ad promoting environmental protection in the same area, she said, ``\textit{[The ad is] very real because it's located in Los Angeles.}'' %(audio1632706610) 
Sometimes, even when users subjectively did not consider an advertisement to be highly related to the persona, they could still speculate about the reasons for encountering the ad. A participant said ``\textit{Max may be tangentially related because he is interested in gaming. So the algorithm might assume that somebody who's interested in gaming might be interested in media services as well.}'' % (audio1854517361, Pos. 1)
\cc{These reasoning processes and emotional reactions demonstrate participants' cognitive and affective empathy toward personas.} Furthermore, it underscores their enhanced privacy knowledge through the examination of ad-persona relationships.
\cc{By summarizing and generalizing the patterns evolved from the correlations, users may leverage such knowledge to reflect on and improve their privacy settings and configurations.}

\textit{(2) Indirect connection based on personal association and stereotypes:} Interestingly, sometimes users were unable to directly pinpoint a specific privacy attribute related to an ad, but still considered the ad relevant to the persona because they made associations based on known privacy information about the persona. Such judgments were sometimes made on their personal biases or societal stereotypes.  A typical example is when a participant believed that a tire ad was related to an African-American persona. Although the persona's profile did not mention any information related to tires, the participant expressed, ``\textit{I think honestly the [ad] save time and money for tires is more for African Americans. Because of African American culture, they like to modify their tires.}'' %(audio1705700564).

These associations based on personal experiences or stereotypes sometimes led participants to draw different conclusions about similar ads. For example, when impersonating the same personas who owned cars, some participants stated that the car advertisement was irrelevant to the persona because "she already has a car (so she wouldn't buy another one)" %(audio1202769299, Pos. 1). 
Others considered the ad as relevant because ``\textit{She might be able to afford to buy a [new] car.}'' 
These association-based inferences, while reflecting participants' cognitive empathy with the persona, raise doubts about whether their judgments help them accurately understand the connection between privacy information and system outcomes.

\section{Discussion}

\subsection{A trade-off in the impact of familiarity and persona realness on empathy (RQ2)}

Previous work in psychology reported that users often exhibit stronger empathy towards individuals they are more familiar with \cite{preston2002empathy, motomura2015interaction}. However, when the object shifts from real people to generated personas, both the user's familiarity with the persona and their perceived realness of the persona impact their empathy towards the persona.

% \tlcomment{I changed the wording from ``realness'' to ``realness'' to be more precise}
% I replaced all the realness with realness to make it consistent

As described in Section \ref{section: familiarity}, there exists a trade-off between the participants' evaluation of the persona's realness and their familiarity with the persona. Specifically, when participants are more familiar with the persona, they are more likely to notice issues within the generated data that make a persona ``appear fake''. This, in turn, results in a negative impact on the user's level of empathy with the persona. While many of these issues are indeed data quality problems, which are more discernible to users deeply versed in the domain, some other ``unrealistic'' details \textit{perceived} by users are linked to biases influenced by the personal experience of users, which we will discuss in Section~\ref{sec:influence_personal_experience}.

Ideally, users should view personas as both familiar and authentic.  Yet, the constraints of current large language models hinder achieving optimal realness \cite{hamalainen2023evaluating}. One method of compensating for the impact of reduced realness is to maintain a balance in familiarity. The goal is to prevent users from seeing personas as too alien or too familiar, which might reveal data inaccuracies. To find the equilibrium and explore the trade-off, we intend to conduct more in-depth user studies, allowing users to experience generated personas with varying levels of realness and familiarity. These studies will help us assess users' empathy towards these personas and delineate the interrelation between familiarity and perceived realness in invoking empathy.

% \hl{[discuss future work opportunity in finding this balance/exploring this trade-off]}
% fixed

% \tlcomment{I changed the wording to ``views'' so it's less ambiguous}

\subsection{The influence of personal views on empathy and privacy literacy (RQ2 \& RQ4)}
\label{sec:influence_personal_experience}
User empathy towards personas is influenced not only by their familiarity with the domain, but also by their personal views. As highlighted in Section~\ref{sec:deviation_experiences}, participants with different personal views perceived the privacy attributes \cc{(e.g., income, weekly schedule, and browsing history)} of the same persona differently. Deviations between a persona's privacy attributes and a user's views can lower the persona's perceived realness. Such views can arise from personal experiences, observations, or even stereotypes. The personal views of users may stem from their own experiences (e.g., a participant working in the finance sector feeling that the data for a ``financial analyst'' persona is not realistic enough), observations of others' experiences (e.g., a participant with research assistant friends believing that research assistants should have longer working hours), or even personal stereotypes (e.g., a participant thinking that African Americans enjoy changing tires). 

These biases can distort the user's understanding of the relationship between privacy attributes and system results, as noted in Section~\ref{sec:perceived_links}. We found that users sometimes explain the results of the system based on personal associations and stereotypes rather than specific privacy attributes, which raises concerns about the accuracy of the knowledge that users acquire. Future work is needed to address the complex interplay between users' personal views, empathy towards personas, and their personal biases stemming from stereotypes. One promising opportunity is to develop sense-making tools to aid users in better understanding and reflecting on their experiences and facilitate the transition of these experiences into accurate privacy knowledge.

Visualization can play a vital role in this direction, assisting users to compare privacy data and system outcomes to foster a more informed and objective understanding of the underlying reasons behind system outcomes. Additionally, research can delve into techniques to mitigate personal biases in user judgments, ensuring that users' interpretations are grounded in objectivity rather than predispositions.

% \hl{[elaborate on the future work opportunity on e.g., sensemaking assistance for users to understand the reasons behind system outcomes; techniques that mitigate personal biases etc.]} added
% \tlcomment{future work needed to help users better make sense of/reflect their experiences/translating their experiences to *correct* knowledge}
% visualization/comparison of privacy data and outcomes 

\subsection{Acquisition of privacy knowledge (RQ3 \& RQ4)}
In Section~\ref{sec:system_sensitivity}, we confirmed that the outcomes of web services and applications are sensitive to privacy modifications by the Privacy Sandbox. Our results also showed that users can indeed perceive the connection between these changes and the persona's privacy data. We expect that this experience will allow them to acquire privacy knowledge and enhance their privacy literacy, which ultimately leads to behavior changes. 

As discussed in Section~\ref{sec:perceived_links}, we observed that users actively engaged emotionally and cognitively when they experienced personalized advertisements associated with the privacy attributes of personas. They felt excited or surprised when they encountered highly relevant advertisements, and were able to identify and internalize the consequences of sharing specific privacy attributes by relating the ads they saw to the privacy attributes of personas. 

This demonstrates the potential of using our proposed approach to support users in experiential learning \cite{gentry1990experiential} for privacy knowledge. Our privacy sandbox prototype provides users with personal involvement in learning privacy knowledge, as both the feelings and the cognitive aspects of users are engaged. Furthermore, this approach serves as a way of \textit{scaffolding} \cite{gonulal2018scaffolding}, enabling users to experience the influence of privacy attributes on system outcomes that they were unable to manage on their own \cite{miller2002theories}. 

Through our approach, users can independently contextualize the privacy attributes of specific personas and understand how the system uses privacy information for personalized content. To validate the effectiveness of this approach in enhancing users' acquisition of privacy knowledge, future steps involve using rigorous tests to assess the privacy knowledge of users before and after using the approach. Furthermore, observing how users apply the privacy knowledge they gain from this approach in simulated or real-world settings can provide deeper insights into how the knowledge impacts users' privacy behaviors.

% \hl{[need to explain why this is scaffolding and what scaffolding is, since the audience of the privacy subcommittee might not be familiar with the concept]} fixed

% \hl{[need to talk about future work in assessing the knowledge acquisition part]} added

\subsection{Biases in LLMs and their impact on generated personas}
Previous work has shown that Large Language Models (LLMs) like GPT risk amplifying existing stereotypes \cite{blodgett2020language, nadeem2020stereoset, smith2022m, hong2023shaping}. However, these biases within the personas generated may not be detrimental in the specific context of our study. 

From the system's perspective, when recommendation systems process personas' privacy data that reflect real-world biases, they will produce representative outcomes resembling the service or experiences in the real world due to the inherited biases stemming from these systems. This means that, even if our generated personas contain biases, they can actually contribute to the realness of the recommendations made by external websites and apps.
% From the user's perspective, the bias within the generated persona data can amplify the bias in system recommendation. Consequently, users may perceive such bias more clearly when they impersonate the privacy personas. Recognizing the systems' bias in handling privacy data benefits users in two folds. It not only assists users in acquiring privacy knowledge but also prompts users to be mindful of the possibility of being inadvertently pigeonholed by the system, resulting in overly tailored and limited services or experiences.

\subsection{Ethical and legal considerations}
While our proposed method has the potential to bring significant benefits to privacy literacy education and positive privacy behavior changes, we also identified some ethical and legal risks in adopting this approach.

\paragraph{Biases may reinforce users' stereotypes.} While the biases and stereotypes in generated privacy personas may make them effective with external recommendation systems, prolonged and repeated exposure to the generated personas with biases may reinforce their pre-existing stereotypes. When using this method, we need to warn users of the risk of potential bias and stereotypes that may be present in the generated persona. Future work is also needed to better mitigate this effect.

\paragraph{Malicious misuse may lead to potential cybersecurity concerns.} While our proposed approach is dedicated to enhancing users' privacy literacy, the method of generating realistic personas to create difficult-to-detect bots or use persona identities for phishing activities. Therefore, this approach necessitates more stringent technical and policy constraints to mitigate potential security concerns associated with its use.

% While our privacy sandbox can be considered useful privacy education tools and digital interventions, we should also consider how it might be adopted in the real world~\cite{10.1145/3370271}.
% For users, there's a risk of reinforcing societal stereotypes. While our study benefited from making these personas recognizable to participants, prolonged and repeated exposure to the generated personas with biases may reinforce users' pre-existing stereotypes. 
% Additionally, our findings indicate that participants sometimes might associate these synthetic personas with real individuals, potentially leading to misrepresentations.
% % Additionally 
% From the system's perspective, as we generate more generated data to train or test systems, these systems are also faced with the increasing challenge of distinguishing between genuine and generated data.  Introducing more generated personas can place additional strains on platforms. 
% It's essential to remain critically aware of the broader ethical implications and potential system strain, deploying them responsibly.

\subsection{Limitation and future work}
We summarize the limitations of our work in four aspects and suggest future steps for each of them: the generation pipeline, the privacy sandbox prototype, the experiment design, and the generalizability of downstream tasks.

% \tlcomment{a tip in writing limitation section -- i usually start with what our work was able to achieve first, before talking about problems.}
\subsubsection{Generation pipeline}
While our current generation pipeline was shown to be generally capable of creating artificial personas that are sufficiently realistic to stimulate changes in system outcomes and invoke user empathy, in Section~\ref{sec:data_consistency}, users pointed out the issues of inconsistencies in the current generated persona data. We found that these inconsistencies were partly due to the inherent randomness embedded in the output of large language models and also resulted from inductive biases \cite{lake2019human} during the few-shot learning process. Such biases can cause generated data to resemble the example data, resulting in users perceiving the generated information as out of context. Our planned future steps to address this issue include fine-tuning the model to enhance data consistency and reduce the generation of irrelevant data. We also plan to store and represent the facts generated about artificial personas more formally in a knowledge graph and improve the coherence within the generated data using knowledge infusion techniques \cite{moiseev2022skill, heinzerling2020language}.

\subsubsection{Privacy sandbox prototype}
The main role of our current Privacy Sandbox prototype was to serve as a proof-of-concept and to support the experiment presented in this paper. Thus, it only supports browser tasks. However, while the web browser is the primary way through which users interact with online services and engage in privacy behaviors, there are other mediums for privacy behaviors that our current privacy sandbox prototype does not support, such as mobile apps and smart home devices. Its support for different types of privacy attributes is also limited, missing support for popular data types such as sensor data and search history. Furthermore, while we have successfully generated realistic social media posts for artificial personas and posted them by invoking the corresponding APIs of social media platforms using scheduled tasks, this approach takes a considerable amount of time (i.e., two weeks to simulate two weeks of social media post history). To address these issues, we will expand our Privacy Sandbox to support other platforms and broaden its support for directly replacing varieties of privacy data types, such as social media posts and sensor data.

\subsubsection{Experiment design}
The goal of the experiment design presented in this paper was to validate the feasibility of our approach in stimulating changes in system outcomes and invoking user empathy. As a result, the experiment did not aim to directly measure the extent to which users acquired privacy knowledge or the subsequent shifts in their privacy behavior as a consequence of our approach. 
Although the qualitative findings of the interviews with participants suggest a strong potential of our approach to achieve these two goals, more work is needed to validate such hypotheses in long term and in-the-wild context. 
\cc{Our study is an initial endeavor to enable users to empathize with generated personas, understand their privacy goals, and gain privacy insights by observing and correlating the system outcomes with privacy attributes.}
In the future, we plan to \cc{collect additional qualitative data when user evaluate the perceived realness across three groups of personas. This will help identify the key factors and issues that influence the perceived realness. We will also} conduct pre-tests and post-tests with users between their usage of our system to assess their acquisition of privacy knowledge, as well as longitudinal deployment studies to measure changes in users' privacy behaviors over time.

% In our study, we did not incorporate an evaluation of ad diversity and the precision of ad recommendations. We also did not directly measure the extent to which users acquired privacy knowledge or the subsequent shifts in their behavior as a consequence of our approach. To address these limitations, we plan to engage experts in recommendation systems to evaluate ad diversity and precision when transitioning between different personas. We will employ pre-tests and post-tests to quantify privacy knowledge acquisition, and we plan to conduct a longitudinal study to assess changes in users' privacy behavior over time.

\subsubsection{Generalizability of downstream tasks} 
\label{sec:downstream tasks}
In our experiment, we chose personalized online advertisements as the target domain of downstream tasks due to their ubiquity, user familiarity, and sensitivity to modification of privacy data. Nevertheless, we expect that our approach can generalize to empowering users to experience the outcomes of their privacy behaviors in a wide range of downstream tasks, such as dynamic social media feeds, algorithmic decision-making, and news recommendations. 
\cc{When acting as the generated personas with various personal profiles and social media posts,  users can explore the association between privacy attributes and new friends or posts on social media. In addition, when browsing the recommended news presented to the generated personas with distinct browsing histories, online interests, and locations, users can also uncover the connections between these attributes with the recommended news.}
For the next phase in the development of our Privacy Sandbox, we plan to add support for these additional downstream tasks, followed by the next rounds of deployment and experiments where we will assess our approach's effectiveness in stimulating empathy, facilitating the acquisition of privacy knowledge, and promoting positive privacy behavior changes in these domains. In the end, upon study results validating its positive impacts, we will publicly release the Privacy Sandbox and promote its adoption through community outreach events for broader impacts.
% only tested our method in the domain of online ads, as we found ads are the most sensitive system outcomes for privacy data modification. 
% However, downstream privacy-sensitive applications can extend beyond ads to areas such as social media, e-commerce, smart home devices, and so on. Additionally, besides enabling users to experience the impact of privacy on system outcomes and acquire privacy knowledge, application developers can also utilize our proposed method to empathize with their target users. This can help them gain a better understanding of what essential privacy information they should request from users, avoid excessive data collection, and summarize a guideline for privacy data usage.

\section{Conclusion}
Aiming to mitigate the privacy gap \cc{between attitudes, goals, knowledge, and behavior,} we introduced an empathy-based approach that allows users to experience how privacy \cc{attributes} may alter system outcomes in a risk-free sandbox environment from the perspective of artificially generated personas. A user study with 15 participants confirmed the quality of the generated personas, validated the effectiveness of our approach in invoking user empathy and system outcome changes, and characterized the impact of users' familiarity, personal experiences, and data consistency on their perceived realness and empathy toward these personas. Our findings offered design implications for implementing this approach in different downstream applications, with the aim of improving user privacy literacy.

\begin{acks}
This work was supported in part by an AnalytiXIN Faculty Fellowship, an NVIDIA Academic Hardware Grant, a Google Cloud Research Credit Award, a Google Research Scholar Award, a Google PSS Privacy Research Award, NSF Grant 2341187, and NSF Grant 2326378. C. Chen and Y. Ye’s work was partially supported by the NSF under grants IIS-2321504, IS-2334193, IIS-2203262, IIS-2217239, CNS-2203261, and CMMI-2146076. Any opinions, findings, and conclusions or recommendations expressed in this material are those of the authors and do not necessarily reflect the views of the sponsors. We would like to thank Tianshi Li and Dakuo Wang for useful discussions.
\end{acks}

\bibliographystyle{ACM-Reference-Format}
\bibliography{biblio}
\appendix

\section{Prompts and few-shot learning examples}
\label{appendix:prompts}

\subsection{Prompts and few-shot learning examples for generating personas using our approach}
\promptbox{
\textbf{Prompt to generate persona description}
\vspace{5pt}

Return a realistic profile. This year is 2023. The income should be in dollars.
The birthday should be in the MM/DD/YYYY format.
The demographic of this person should represent the US population sample.

The generated profile should match the following guidance: <\textcolor{red}{guidance}>.

Fit into the braces in the profile:\\
\{\textcolor{brown}{First name}\} \{\textcolor{brown}{Last name}\} is a \{\textcolor{brown}{age ranging from 18 to 70 subject to continuous uniform distribution}\} \{\textcolor{brown}{race}\} \{\textcolor{brown}{gender}\} 
living in \{\textcolor{brown}{real home address with street, city, state, and zip code}\}. \{\textcolor{brown}{Pronoun}\} speaks \{\textcolor{brown}{spoken language}\}. {\textcolor{brown}{Pronoun}}'s education background is \{\textcolor{brown}{educational background}\}. \{\textcolor{brown}{Pronoun}\}'s date of birth is \{\textcolor{brown}{date of birth}\}. \{\textcolor{brown}{Pronoun}\} is a \{\textcolor{brown}{occupation}\}, and the annual income is \{\textcolor{brown}{income in dollar}\}. \{\textcolor{brown}{marital status}\} \{\textcolor{brown}{parental status}\} \{\textcolor{brown}{detailed habits and preferences when using the computer, mobile phone, and the Internet}\}.\\

The format of the generated result should look like the following examples: <\textcolor{blue}{few-shot learning example}>

Return the profile in only one paragraph.
}

\promptbox{
\textbf{Few-shot learning example for the generation of persona description}
\vspace{5pt}

\textcolor{blue}{Abigail Patel is a 32-year-old Asian American female living at 325 Main St, Newark, NJ 07102. She speaks English and her educational background includes a bachelor's degree in Marketing. Abigail's date of birth is 05/26/1991. She is currently working as a marketing manager, with an annual income of \$85,000. Abigail is married and has two children. She enjoys browsing social media and streaming movies on her mobile phone during her free time. When using her computer, she prefers using a wireless mouse and keyboard for easy navigation. On the internet, she likes to shop for clothes and read reviews before making a purchase.}
}

\promptbox{
\textbf{Prompt to generate privacy attributes}
\vspace{5pt}

<\textcolor{blue}{few-shot learning example}>

Given the profile: <persona>.\\
Return the attributes in this format:\\
\textcolor{brown}{\{``first name'': ``'', ``last name'': ``'', ``age'': ``'', ``gender'': ``'', ``race'': ``'', ``street'': ``'', ``city'': ``'', ``state'': ``'', "zip code'': ``'', ``spoken language'': ``'', ``educational background'': ``'', ``birthday'': ``'', ``job'': ``'', ``income'': ``'', ``marital status'': ``'', ``parental status'': ``'', ``online behavior'': ``''\}}
}

\promptbox{
\textbf{Few-shot learning example for the generation of privacy attributes}
\vspace{5pt}

\textcolor{blue}{Given the profile: Abigail Patel is a 32-year-old Asian American female living at 325 Main St, Newark, NJ 07102. She speaks English and her educational background includes a bachelor's degree in Marketing. Abigail's date of birth is 05/26/1991. She is currently working as a marketing manager, with an annual income of \$85,000. Abigail is married and has two children. She enjoys browsing social media and streaming movies on her mobile phone during her free time. When using her computer, she prefers using a wireless mouse and keyboard for easy navigation. On the internet, she likes to shop for clothes and read reviews before making a purchase.\\
Return the attributes in this format:\\
\{``first name'': ``Abigail'', ``last name'': ``Patel'', ``age'': ``32'', ``gender'': ``female'', ``race'': ``Asian American'', ``street'': ``325 Main St'', ``city'': ``Newark'', ``state'': ``NJ'', "zip code'': ``07102'', ``spoken language'': ``English'', ``educational background'': ``bachelor's degree in Marketing'', ``birthday'': ``05/26/1991'', ``job'': ``marketing manager'', ``income'': ``85,000'', ``marital status'': ``married'', ``parental status'': ``has two children'', ``online behavior'': ``She enjoys browsing social media and streaming movies on her mobile phone during her free time. When using her computer, she prefers using a wireless mouse and keyboard for easy navigation. On the internet, she likes to shop for clothes and read reviews before making a purchase.''\}}
}

\promptbox{
\textbf{Prompt for generating portrait image prompt}
\vspace{5pt}

Given the profile description, output a descriptive prompt to generate a realistic human head portrait image, limit the word to 30 words: \{\textcolor{brown}{persona description}\}
}

\promptbox{
\textbf{Prompt to generate device and browser}
\vspace{5pt}

Given the profile: \{\textcolor{brown}{persona}\}, infer the browser and device the person uses:
}

\promptbox{
\textbf{Prompt to generate schedule}
\vspace{5pt}

<\textcolor{blue}{few-shot learning example}>\\
You are acting as a game event designer.
Write daily events for this person: \{\textcolor{red}{persona description}\}. 
Show me a reasonable schedule for this person from \{\textcolor{red}{start\_date}\} to \{\textcolor{red}{end\_date}\}.
The life in the period is similar to 2021.
You can generate fake but reasonable data that is related to the profile.
The start time of one day is 00:00:00.
Generate events from 00:00:00 to 23:59:59 for each day.\\
Return a list of dict.\\
Output the following JSON format in plain text:\\
\textcolor{brown}{
\{
    ``start\ time'': <start moment of the event>,
    ``end\ time'': <start moment of the event>,
    ``event'': <event>
\}}\\
Never provide additional context.
}

\promptbox{
\textbf{Few-shot learning example for the generation of schedule}
\vspace{5pt}

\textcolor{blue}{
``profile'': ``Daniel Chan is a 30\-year\-old Asian man living in Seattle, Washington with zip code 98101. He is a project manager, and the annual income is around one hundred and twenty thousand USD. He lives alone in a studio apartment and likes to keep his space clean and organized. In his free time, he enjoys playing video games and reading books on his Kindle. He also likes to use social media platforms such as Twitter and Reddit to keep up with the latest news and trends. ''\\
``location\_history'':
        [
            [``2023-06-05 00:00:00'', ``2023-06-05 07:00:00'', ``Home - 1420 5th Ave, Seattle, WA 98101''],
            [``2023-06-05 07:00:00'', ``2023-06-05 08:30:00'', ``Golds Gym - 1220 Howell St, Seattle, WA 98101''],
            [``2023-06-05 08:30:00'', ``2023-06-05 09:00:00'', ``Starbucks - 1125 4th Ave, Seattle, WA 98101'']
        ]
}
}

\promptbox{
\textbf{Prompt to generate browsing history}
\vspace{5pt}

<\textcolor{blue}{few-shot learning example}>\\
Given the person's profile: \{\textcolor{red}{persona description}\}, 
and the schedule: \{\textcolor{red}{schedule}\},
generate \{\textcolor{red}{number}\} browser history entries from \{\textcolor{red}{start\_date}\} to \{\textcolor{red}{end\_date}\}.\\
No browsing history between 00:00:00 and 07:00:00.
The webpage title should reflect the content in the webpage url.
The webpage be reasonable and related to the the schedule.
Don't add the address of the schedule to the webpage title.
The datetime should be realistic and associated with the webpage content.
The datetime second should not be 0.
The datetime should be dispensed in one day.\\
You can generate fake but reasonable data that is consistent with the profile and schedule.
Output following list format in plain text:\textcolor{brown}{[[<datetime>, <webpage titile>, <webpage url>],]}\\
Never provide additional context.
}

\promptbox{
\textbf{Few-shot learning example for the generation of browser history}
\vspace{5pt}

\textcolor{blue}{
``profile'': ``John Smith is a 25\-year\-old Caucasian male living at 123 Park Ave, New York, NY 10001. He speaks English and his educational background includes studying Computer Science and Data Analysis. John's date of birth is 09/15/1998. He is currently a student, and his annual income is \$5000. John is single and does not have any children. He enjoys coding and exploring new technologies on his computer. On his mobile phone, he prefers using apps for productivity and staying up to date with the latest tech news. When using the Internet, he enjoys participating in online coding forums and watching tutorial videos to enhance his skills.''\\
\\
``schedule'':[
    [`2023-07-10 00:00:00', `2023-07-10 07:00:00', `Home - 123 Park Ave, New York], 
    [`2023-07-10 07:00:00', `2023-07-10 08:00:00', `Morning Exercise - 987 8th Ave, New York, NY 10019'], 
    [`2023-07-10 08:00:00', `2023-07-10 09:00:00', `Breakfast - 654 Hudson St, New York, NY 10014'], 
    [`2023-07-10 09:00:00', `2023-07-10 12:00:00', `Study Computer Science - 101 Lafayette St, New York, NY 10013'], 
    [`2023-07-10 12:00:00', `2023-07-10 13:00:00', `Lunch - 246 Spring St, New York, NY 10013'], 
    [`2023-07-10 13:00:00', `2023-07-10 15:00:00', `Online Coding Forums - 876 4th Ave, New York, NY 10018'], 
    [`2023-07-10 15:00:00', `2023-07-10 17:00:00', `Study Data Analysis - 321 Canal St, New York, NY 10013'], 
    [`2023-07-10 17:00:00', `2023-07-10 18:00:00', `Break - 789 6th Ave, New York, NY 10001'], 
    [`2023-07-10 18:00:00', `2023-07-10 20:00:00', `Dinner - 897 Broadway, New York, NY 10003'], 
    [`2023-07-10 20:00:00', `2023-07-10 23:59:59', `Free Time - 456 Broadway, New York, NY 10013']
],\\     
\\
"browser\_history":[
    [`2023-07-10 15:27:08', `Learning Log: Consider how data analysts approach tasks', \\`https://www.coursera.org/learn/foundations-data/supplement/I086K/learning-log-consider-how-data-analysts-approach-tasks'],
    [`2023-07-10 15:24:41', `Case Study: New data perspectives', \\`https://www.coursera.org/learn/foundations-data/supplement/nhC19/case-study-new-data-perspectives'],
    [`2023-07-10 15:21:14', `Data analytics in everyday life', 
    \\`https://www.coursera.org/learn/foundations-data/lecture/N5lvQ/data-analytics-in-everyday-life']
 ]
}
}

\promptbox{
\textbf{Prompt to generate social media post content}
\vspace{5pt}

Provide ideas for this person to write posts (limit
the word to 140 words) based on the profile and location history: \{\textcolor{brown}{profile}\} \{\textcolor{brown}{schedule}\}

The schedule is in the format of [[start time, end time, address]].

Show me only {num} reasonable description in total between {start\_date} and {end\_date} to provide ideas.\
The life in the given time period is similar to 2021 so you can generate the description based on your current data.\

You should only return the list to me without any explanation message. You don't need to use any real-time data, just generate reasonable and consistent data.
You don't need to generate descriptions that may be inappropriate, irrelevant, or offensive.
You do not need to manipulate the data in a way that is specific to a given time period.
The seconds in the time should not be 00, it should be the format like 15:23:12.

Output the following JSON format in plain text:
\textcolor{brown}{
[\{
    "time": <time in string format>,
    "address": <address where this person share the life>,
    "content": <content>,
\}]
}\\
Never provide additional context.
}

\promptbox{
\textbf{Few-shot learning example for the generation of social media post content}
\vspace{5pt}

\textcolor{blue}{
``profile'': ``Emily Rodriguez is a 46\-year\-old Hispanic female living in 602 S Fairfax Ave, Los Angeles, CA 90036. She works as a nurse and earns an annual income of \$70,000. Emily is happily married with two children who are currently in college. In her free time, she enjoys reading and gardening. Emily prefers using her mobile phone for browsing social media and checking emails while using her laptop for work-related tasks. She is mindful of her online security and regularly updates her passwords and privacy settings.''\\
\\
``schedule'':[[`2023-06-06 00:00:00', `2023-06-06 07:30:00', `Home - 123 Main St, Los Angeles, CA 90022'], [`2023-06-06 07:30:00', `2023-06-06 08:15:00', `Starbucks - 5353 E Olympic Blvd, Los Angeles, CA 90022'], [`2023-06-06 08:15:00', `2023-06-06 12:00:00', `Tech Office - 3000 E 1st St, Los Angeles, CA 90063'], [`2023-06-06 12:00:00', `2023-06-06 12:45:00', `Lunch Spot - 3000 E 1st St, Los Angeles, CA 90063'], [`2023-06-06 12:45:00', `2023-06-06 17:30:00', `Tech Office - 3000 E 1st St, Los Angeles, CA 90063'], [`2023-06-06 17:30:00', `2023-06-06 19:00:00', `Gym - 1234 Whittier Blvd, Los Angeles, CA 90022'], [`2023-06-06 19:00:00', `2023-06-06 19:45:00', `Grocery Store - 5432 Whittier Blvd, Los Angeles, CA 90022'], [`2023-06-06 19:45:00', `2023-06-06 21:00:00', `Home - 123 Main St, Los Angeles, CA 90022'], [`2023-06-06 21:00:00', `2023-06-06 22:30:00", `Favorite Local Park - 5432 E 4th St, Los Angeles, CA 90022'], [`2023-06-06 22:30:00', `2023-06-06 23:59:59", `Home - 123 Main St, Los Angeles, CA 90022']
],\\     
\\
"posts": [[`2023-06-06 07:31:42', `Starting my day with a refreshing cup of coffee at Starbucks. Ready to tackle another day at work! \#CoffeeLover \#WorkLifeBalance', `Starbucks - 5353 E Olympic Blvd, Los Angeles, CA 90022'], [`2023-06-06 19:01:02', `Feeling the burn at the gym! Taking care of my health and fitness is a top priority. \#FitnessJourney \#HealthyLiving', `Gym - 1234 Whittier Blvd, Los Angeles, CA 90022']]}
}

\promptbox{
\textbf{Prompt for the prompt to generate social media post image}
\vspace{5pt}

Given the post \{\textcolor{brown}{content}\}, output a descriptive prompt to generate a realistic life image, limit
the word to 30 words:
}

\subsection{Prompts for generating personas using baseline GPT}
\promptbox{
\textbf{Prompt to generate persona description}
\vspace{5pt}

Return a realistic profile. This year is 2023. The income should be in dollars.
The birthday should be in the MM/DD/YYYY format.
The demographic of this person should represent the US population sample.

The generated profile should match the following guidance: <\textcolor{red}{guidance}>.

Fit into the braces in the profile:\\
\{\textcolor{brown}{First name}\} \{\textcolor{brown}{Last name}\} is a \{\textcolor{brown}{age ranging from 18 to 70 subject to continuous uniform distribution}\} \{\textcolor{brown}{race}\} \{\textcolor{brown}{gender}\} 
living in \{\textcolor{brown}{real home address with street, city, state, and zip code}\}. \{\textcolor{brown}{Pronoun}\} speaks \{\textcolor{brown}{spoken language}\}. {\textcolor{brown}{Pronoun}}'s education background is \{\textcolor{brown}{educational background}\}. \{\textcolor{brown}{Pronoun}\}'s date of birth is \{\textcolor{brown}{date of birth}\}. \{\textcolor{brown}{Pronoun}\} is a \{\textcolor{brown}{occupation}\}, and the annual income is \{\textcolor{brown}{income in dollar}\}. \{\textcolor{brown}{marital status}\} \{\textcolor{brown}{parental status}\} \{\textcolor{brown}{detailed habits and preferences when using the computer, mobile phone, and the Internet}\}.\\

Return the profile in only one paragraph.
}

\promptbox{
\textbf{Prompt to generate privacy attributes}
\vspace{5pt}

Given the profile: <persona>.\\
Return the attributes in this format:\\
\textcolor{brown}{\{``first name'': ``'', ``last name'': ``'', ``age'': ``'', ``gender'': ``'', ``race'': ``'', ``street'': ``'', ``city'': ``'', ``state'': ``'', "zip code'': ``'', ``spoken language'': ``'', ``educational background'': ``'', ``birthday'': ``'', ``job'': ``'', ``income'': ``'', ``marital status'': ``'', ``parental status'': ``'', ``online behavior'': ``''\}}
}

\promptbox{
\textbf{Prompt for generating portrait image}
\vspace{5pt}

Generate a realistic human head portrait image
}

\promptbox{
\textbf{Prompt to generate device and browser}
\vspace{5pt}

Generate the browser and device a person uses:
}

\promptbox{

\textbf{Prompt to generate schedule}
\vspace{5pt}

You are acting as a game event designer.
Write daily events for a persona. 
Show me a reasonable schedule for this person from \{\textcolor{red}{start\_date}\} to \{\textcolor{red}{end\_date}\}.
The life in the period is similar to 2021.
You can generate fake but reasonable data that is related to the profile.
The start time of one day is 00:00:00.
Generate events from 00:00:00 to 23:59:59 for each day.\\
Return a list of dict.\\
Output the following JSON format in plain text:\\
\textcolor{brown}{
\{
    ``start\ time'': <start moment of the event>,
    ``end\ time'': <start moment of the event>,
    ``event'': <event>
\}}\\
Never provide additional context.
}

\promptbox{

\textbf{Prompt to generate browsing history}
\vspace{5pt}

Generate \{\textcolor{red}{number}\} browser history entries from \{\textcolor{red}{start\_date}\} to \{\textcolor{red}{end\_date}\}.\\
No browsing history between 00:00:00 and 07:00:00.
The webpage title should reflect the content in the webpage url.
The webpage be reasonable and related to the the schedule.
Don't add the address of the schedule to the webpage title.
The datetime should be realistic and associated with the webpage content.
The datetime second should not be 0.
The datetime should be dispensed in one day.\\
You can generate fake but reasonable data that is consistent with the profile and schedule.
Output following list format in plain text:\textcolor{brown}{[[<datetime>, <webpage titile>, <webpage url>],]}\\
Never provide additional context.
}

\promptbox{

\textbf{Prompt to generate social media post content}
\vspace{5pt}

Provide ideas for a person to write posts (limit
the word to 140 words) 

Show me only {num} reasonable description in total between {start\_date} and {end\_date} to provide ideas.\
The life in the given time period is similar to 2021 so you can generate the description based on your current data.\

You should only return the list to me without any explanation message. You don't need to use any real-time data, just generate reasonable and consistent data.
You don't need to generate descriptions that may be inappropriate, irrelevant, or offensive.
You do not need to manipulate the data in a way that is specific to a given time period.
The seconds in the time should not be 00, it should be the format like 15:23:12.

Output the following JSON format in plain text:
\textcolor{brown}{
[\{
    "time": <time in string format>,
    "address": <address where this person share the life>,
    "content": <content>,
\}]
}\\
Never provide additional context.
}

\promptbox{
\textbf{Prompt to generate social media post image}
\vspace{5pt}

Generate a realistic life image for social media posts
}
\section{Participants' demographic data}
\label{appendix:participant_demographic}

\begin{table}[h]
\small
\centering
\begin{tabular}{ccccc}
\toprule
Persona ID & Age & Gender & Education level & Digital Literacy\\
\midrule
1 & 26 & Female     & Master's      & 5 \\
2 & 26 & Male       & Ph.D.         & 5 \\
3 & 25 & Female     & Bachelor's    & 5 \\
4 & 25 & Male       & Bachelor's    & 3\\
5 & 28 & Female     & Master's      & 5\\
6 & 25 & Female     & Master's      & 4\\
7 & 24 & Female     & Master's      & 5\\
8 & 25 & Female     & Master's      & 4\\
9 & 20 & Male       & High school Diploma & 5\\
10 & 33 & Male      & Ph.D.         & 5\\
11 & 26 & Male      & Master's      & 5\\
12 & 20 & Male      & High school Diploma & 2\\
13 & 19 & Female    & High school Diploma & 4\\
14 & 23 & Female    & Master's      & 4\\
15 & 28 & Female    & Master's      & 4\\
\bottomrule
\end{tabular}
\caption{The demographic information of participants (The digital literacy is self rated)}
\label{tab: participant_demographic}
\end{table}
\section{CODE SYSTEM 1}
\label{appendix:codebook}

This is the code system according to which the qualitative data from part one of the interview has been coded.

\subsection{Codes}

\begin{enumerate}
    \item Emotional responses
    \begin{enumerate}
        \item Ads
        \begin{enumerate}
        \item Feel confused when failing to understand the ad
        \item Feel excited that the ad targeting is accurate to the persona's information
        \end{enumerate}
        \item personal information
        \begin{enumerate}
               \item Feel bored with uninteresting photos in the posts
        \item Feel suspicious toward overly consistent posts and events
        \item Feel confused about who takes the photos in the posts
        \item Feel satisfied \& surprised when seeing diverse browsing history
        \item Feel confused about distances between different places
        \item Have a higher tolerance for persona in unfamiliar fields
        \item Feel excited when observing consistent information
        \end{enumerate}
    
    \end{enumerate}

       \item Type of Non-authenticity

    \begin{enumerate}
        \item Inconsistent with personal experience
        \begin{enumerate}
         
        \item Personal description
        \begin{enumerate}
            \item income is too low
        \end{enumerate}

            \item Social media is not real
            \begin{enumerate}
                \item Revealing personal privacy
                \item Sharing only factual content without emotions and thoughts
                \item Repetitive posts across platforms
              
                \item Too frequent and repetitive posting
            \end{enumerate}
            \item Schedule and Time Management 
              \begin{enumerate}
                \item Over-organized schedule
                \item Unreasonable time allocations (e.g., too short work time, too long exercise time)
                \item Schedule being too tight
                \item Inconsistent work intervals
                \item Too few occasional events
            \end{enumerate}
             \item Lifestyle
              \begin{enumerate}
                \item Excessive or insufficient grocery shopping
                \item Too few occasional incidents or events
                \item Inconsistent income with the lifestyle presented
                \item Excessive exercise or lack thereof
                \item Inconsistent times for activities (e.g., too early gym sessions)
            \end{enumerate}
               
        \end{enumerate}
        \item Consistency of Information
        \begin{enumerate}
                \item Picture does not match personal info
            \item Interest is inconsistent with education 
            \item Background posts are inconsistent with events 
            \item Event is inconsistent with personal description
            \item Inconsistent title for same link
            \item Events are not consistent with profile
            \item Browsing history is inconsistent with events
            \item Browsing history should be more consistent with hobbies
            \item URL is inconsistent with title picture does not match income
            \item Inconsistent events
            \item Income does not match with job 
            \item Inconsistent location
            \item Picture does not match with content 
            \item Inconsistency between picture
            \item Browsing history
            \begin{enumerate}
                \item Browsing history appearing at unlikely times
                \item Repetitive browsing behavior
                \item Identical timestamps in browsing history
                \item Browsing sites too basic for an experienced person
                \item No connection between browsing history and personal info
                \item Incorrect link title
                \item Lack of record in certain time period
                \item Ssome events should not have browsing history every day
            \end{enumerate}

            \item Out of context issues
            \begin{enumerate}
                \item Content lacking connection to personal info
                \item Missing specificity and details in content
            \end{enumerate}
        \end{enumerate}
    \end{enumerate} 

    \item Privacy attributes contributing to authenticity
    \begin{enumerate}
        \item Consistency
        \begin{enumerate}
            \item Events are consistent with the job
            \item With personal experience
            \item Events are consistent with hobbies
            \item Social media posts are consistent with weekly schedule
            \item Browsing history is consistent with hobbies
            \item Events among a week are consistent
        \end{enumerate}
    \end{enumerate}
    
    \item Strategies to modify information
    \begin{enumerate}
        \item Be more specific
        \item Increase/decrease certain activities
        \item Modify work and leisure schedules
        \item Browsing in a progressive way
        \item Some events can be more dispersed
        \item Increase salary
        \item Express more diverse emotion and attitude
        \item More life-oriented browsing history
    \end{enumerate}
     
    \item Reasons for irrelevant ads
    \begin{enumerate}
        \item Not mentioned in the persona's data
        \item Inconsistent with persona's data 
         \begin{enumerate}
        \item Profile photos
        \item Race
        \item Social media post
        \item Location
        \item Marital status
        \item Job
        \item Interest or hobbies
        \item Income
        \item Age and state
        \item daily activities
        \item Personal property
        \item Gender
    \end{enumerate}
        \item Looks like spam
        \item Already have or know similar things
        \item Ad feels too generic
        \item Choose specific websites instead of ad
    \end{enumerate}
    
    \item Perceived privacy attributes affecting ads
    \begin{enumerate}
        \item Gender
        \item Social media post
        \item Personal property
        \item Income
        \item Parental status
        \item Job related location
        \item Daily activities
        \item Marital status
        \item Interests and hobbies
        \item Race
        \item Age and state
    \end{enumerate}
\end{enumerate}

% \subsection{Code Definitions and Examples}
% \begin{enumerate}
%     \item \textbf{Type of Non-authenticity}: Refers to the uniformity and coherence of the information provided. 
%     \begin{enumerate}
%     \item \textbf{Consistency of Information}. Refers to the uniformity and coherence of the information provided.\textit{Examples:}
%        \item \textbf{Authenticity of Details}.  Refers to the genuineness of specific details.\textit{Examples:}
%           \item \textbf{Randomness}. Refers to the presence of arbitrary or inconsistent details.\textit{Examples:}

%     \end{enumerate}
    
% \end{enumerate}
\section{Personas generated by our approach}
\label{appendix:our_personas}

\begin{figure*}
    \centering
    \includegraphics[width=\linewidth]{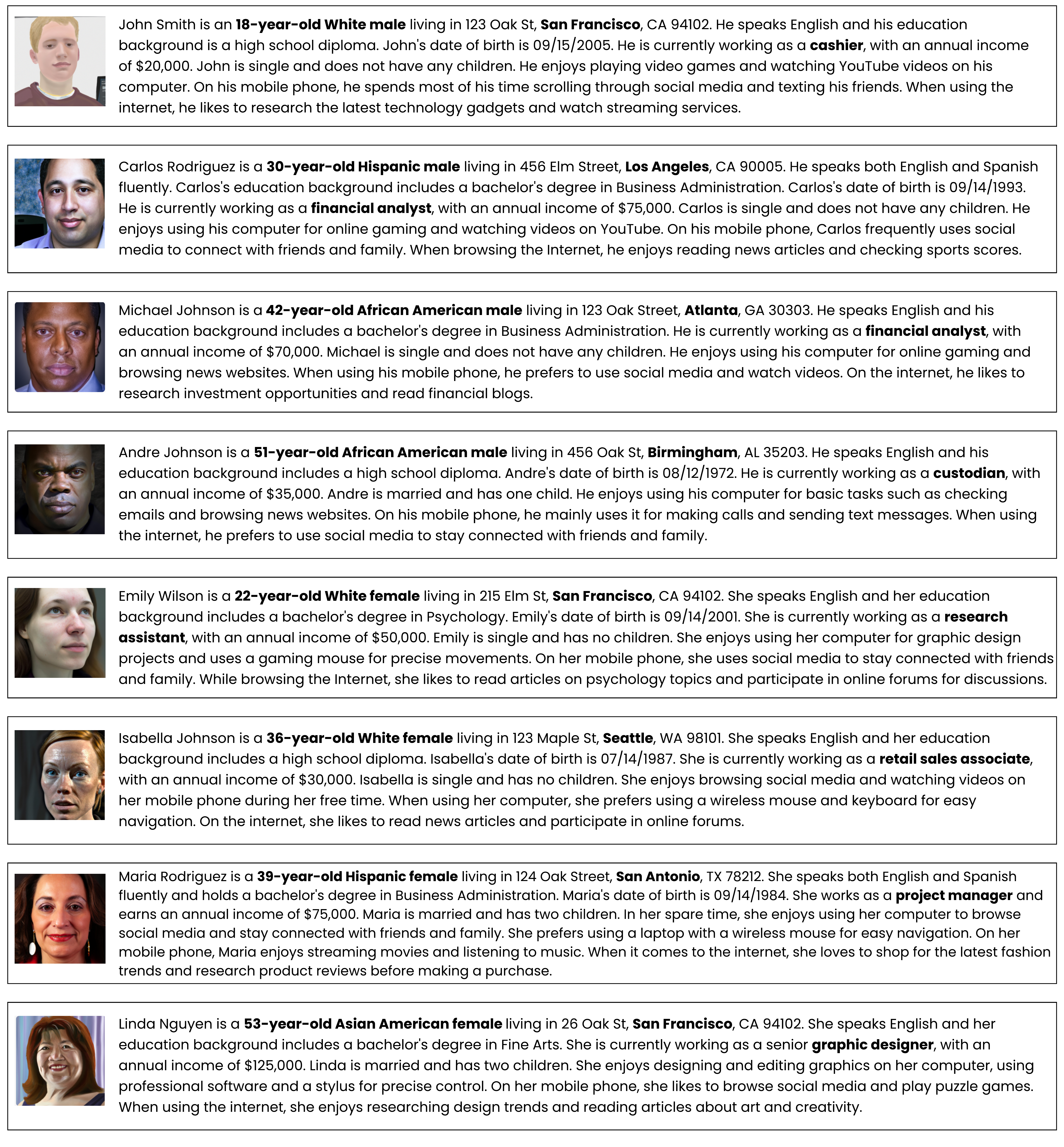}
    \caption{Personas generated by our approach.}
\end{figure*}

\begin{figure*}
    \centering
    \includegraphics[width=\linewidth]{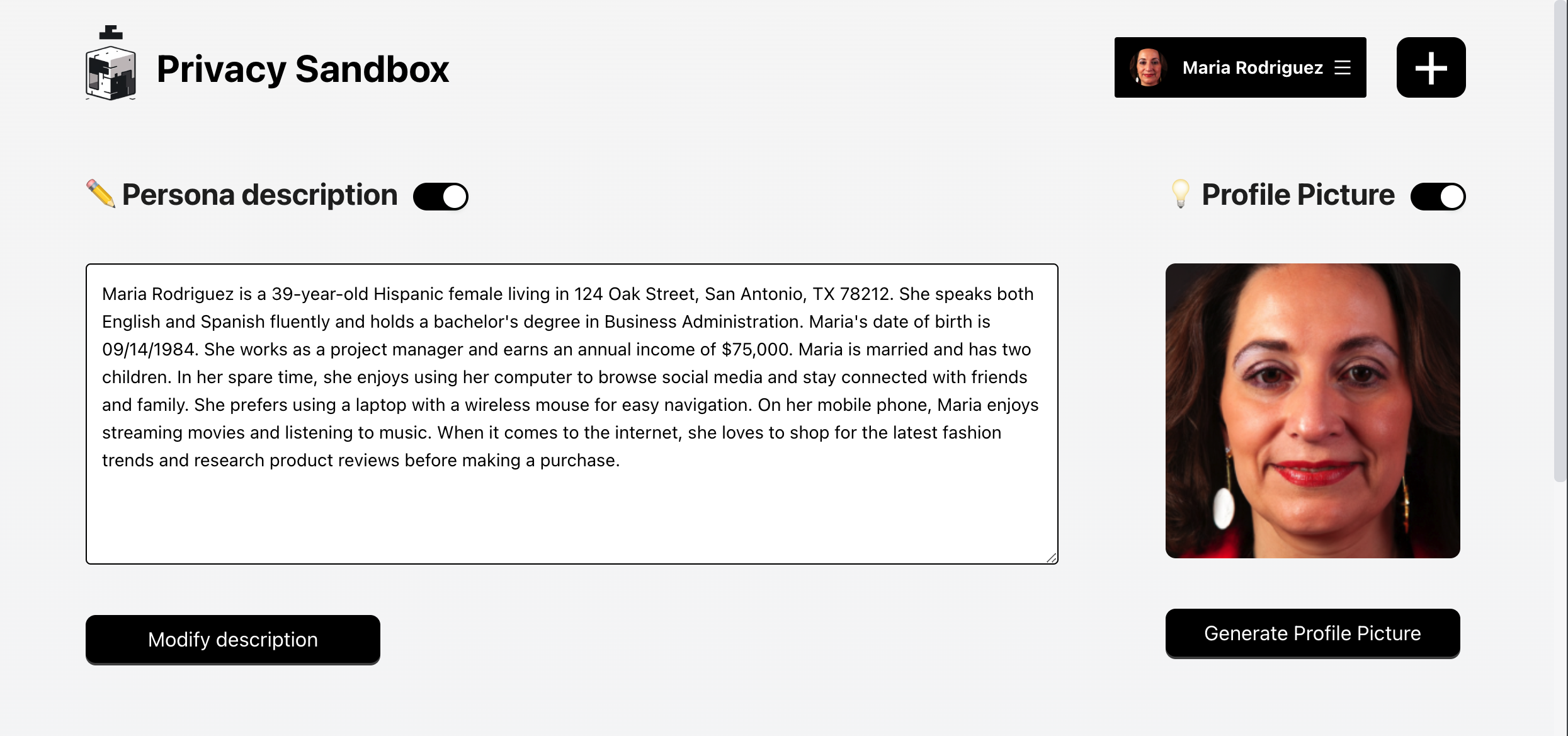}
    \caption{Screenshot of persona description of Maria}
\end{figure*}

\begin{figure*}
    \centering
    \includegraphics[width=\linewidth]{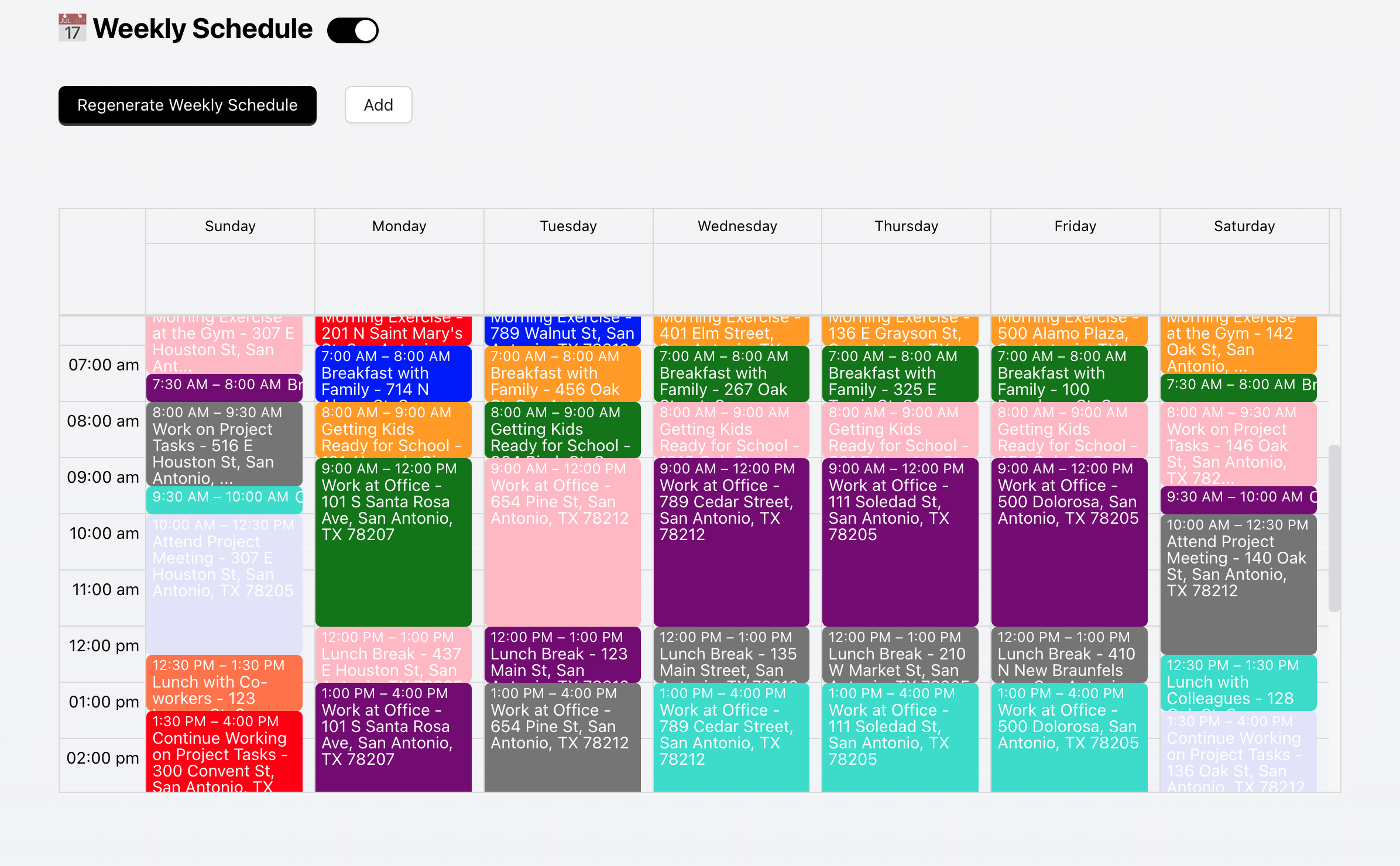}
    \caption{Screenshot of schedule of Maria}
\end{figure*}

\begin{figure*}
    \centering
    \includegraphics[width=0.72\linewidth]{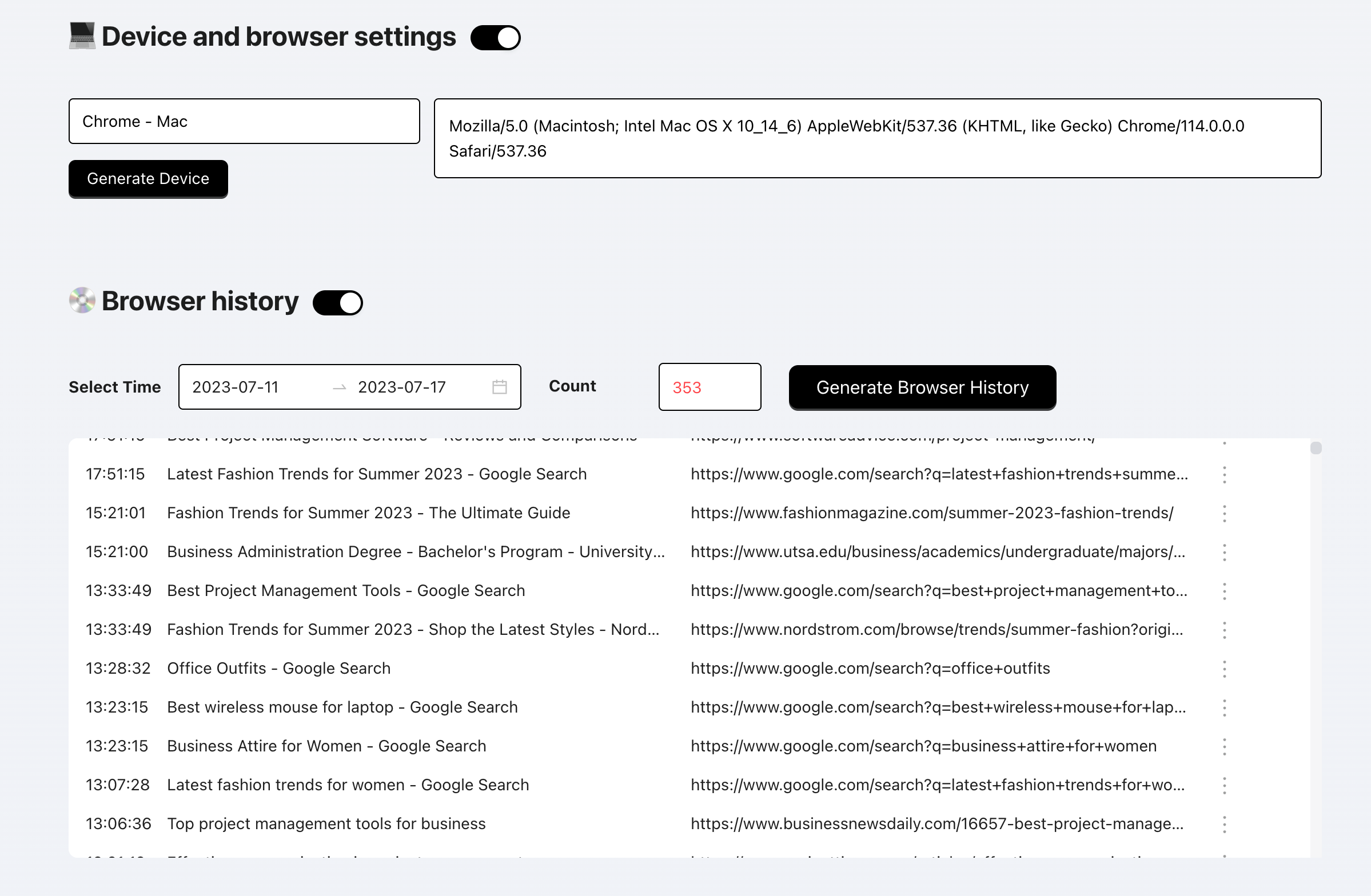}
    \caption{Screenshot of device and browsing history of Maria}
\end{figure*}

\begin{figure*}
    \centering
    \includegraphics[width=0.72\linewidth]{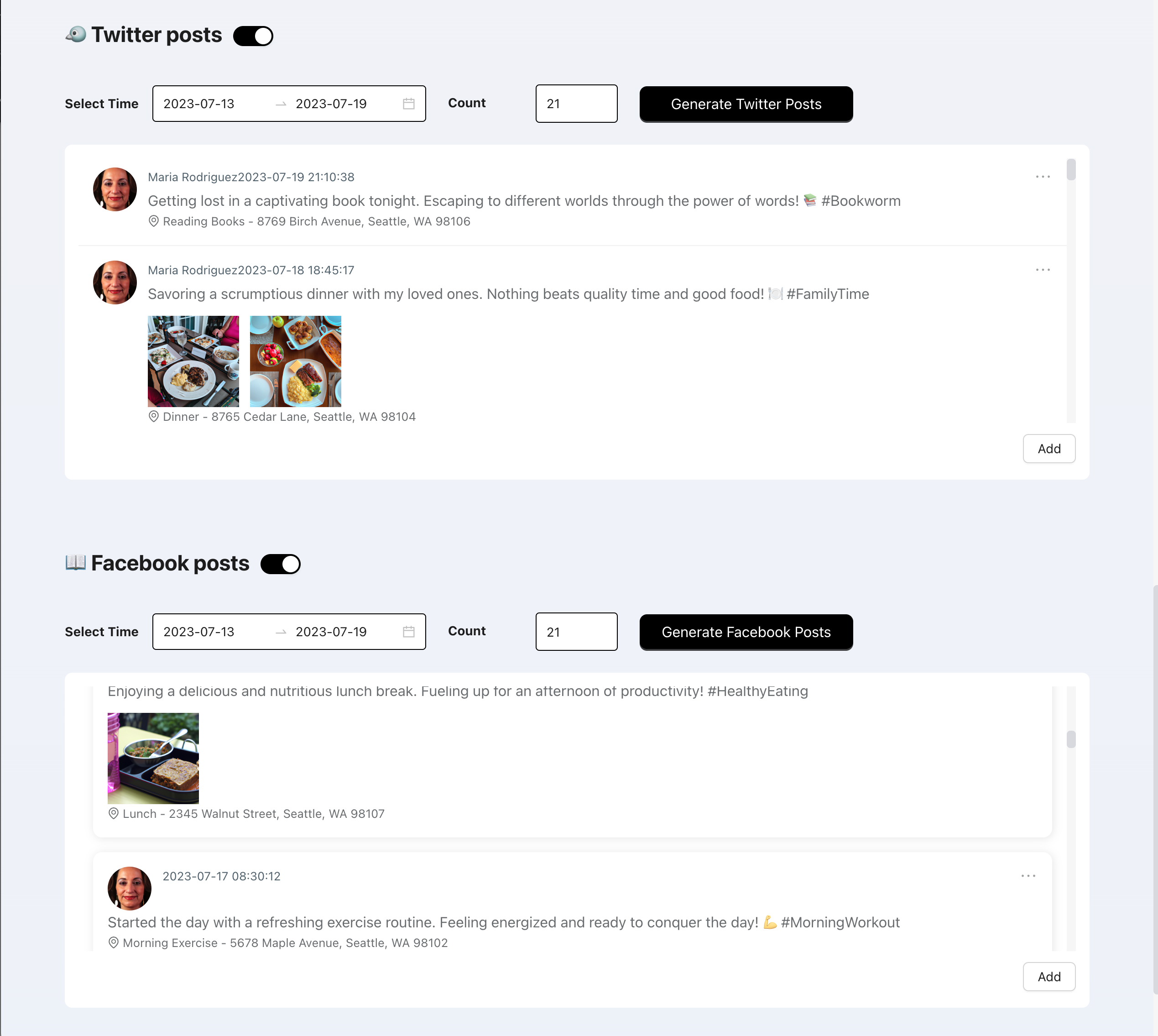}
    \caption{Screenshot of social posts of Maria}
\end{figure*}

\end{document}